\documentclass[fleqn,usenatbib]{mnras}

\usepackage{newtxtext,newtxmath}

\usepackage[T1]{fontenc}
\usepackage{ae,aecompl}

\usepackage{graphicx}	
\usepackage{amsmath}	
\usepackage{amssymb}	
\usepackage{rotating}
\usepackage{color} 
\usepackage{outlines}
\usepackage{todonotes}
\usepackage{lineno}

\usepackage{eso-pic}

\AddToShipoutPictureBG*{%
  \AtPageUpperLeft{%
    \hspace{0.75\paperwidth}%
    \raisebox{-3.5\baselineskip}{%
      \makebox[0pt][l]{\textnormal{DES 2020-0535}}
}}}%

\AddToShipoutPictureBG*{%
  \AtPageUpperLeft{%
    \hspace{0.75\paperwidth}%
    \raisebox{-4.5\baselineskip}{%
      \makebox[0pt][l]{\textnormal{Fermilab PUB-20-143-E}}
}}}%

\title[Undetected sources in DES]{Noise from Undetected Sources in Dark Energy Survey Images}

\author[DES Collaboration]{
\parbox{\textwidth}{
\Large
K.~Eckert,$^{1}$\thanks{E-mail: keckert@sas.upenn.edu}
G.~M.~Bernstein,$^{1}$\thanks{E-mail: garyb@physics.upenn.edu}
A.~Amara,$^{2}$
A.~Amon,$^{3}$
A.~Choi,$^{4}$
S.~Everett,$^{5}$
D.~Gruen,$^{6,3,7}$
R.~A.~Gruendl,$^{8,9}$
E.~M.~Huff,$^{10}$
N.~Kuropatkin,$^{11}$
A.~Roodman,$^{3,7}$
E.~Sheldon,$^{12}$
B.~Yanny,$^{11}$
Y.~Zhang,$^{11}$
T.~M.~C.~Abbott,$^{13}$
M.~Aguena,$^{14,15}$
S.~Avila,$^{16}$
K.~Bechtol,$^{17,18}$
D.~Brooks,$^{19}$
D.~L.~Burke,$^{3,7}$
A.~Carnero~Rosell,$^{20}$
M.~Carrasco~Kind,$^{8,9}$
J.~Carretero,$^{21}$
M.~Costanzi,$^{22,23}$
L.~N.~da Costa,$^{15,24}$
J.~De~Vicente,$^{25}$
S.~Desai,$^{26}$
H.~T.~Diehl,$^{11}$
J.~P.~Dietrich,$^{27}$
T.~F.~Eifler,$^{28,10}$
A.~E.~Evrard,$^{29,30}$
B.~Flaugher,$^{11}$
J.~Frieman,$^{11,31}$
J.~Garc\'ia-Bellido,$^{16}$
E.~Gaztanaga,$^{32,33}$
J.~Gschwend,$^{15,24}$
G.~Gutierrez,$^{11}$
W.~G.~Hartley,$^{48,19,2}$
D.~L.~Hollowood,$^{5}$
K.~Honscheid,$^{4,34}$
D.~J.~James,$^{35}$
R.~Kron,$^{11,31}$
K.~Kuehn,$^{36,37}$
M.~A.~G.~Maia,$^{15,24}$
J.~L.~Marshall,$^{38}$
P.~Melchior,$^{39}$
F.~Menanteau,$^{8,9}$
R.~Miquel,$^{40,21}$
R.~L.~C.~Ogando,$^{15,24}$
A.~Palmese,$^{11,31}$
F.~Paz-Chinch\'{o}n,$^{49,9}$
A.~A.~Plazas,$^{39}$
A.~K.~Romer,$^{41}$
E.~Sanchez,$^{25}$
V.~Scarpine,$^{11}$
S.~Serrano,$^{32,33}$
I.~Sevilla-Noarbe,$^{25}$
M.~Smith,$^{42}$
M.~Soares-Santos$^{50}$
 E.~Suchyta,$^{43}$
M.~E.~C.~Swanson,$^{9}$
G.~Tarle,$^{30}$
D.~Thomas,$^{44}$
T.~N.~Varga,$^{45,46}$
A.~R.~Walker,$^{13}$
W.~Wester,$^{11}$
R.D.~Wilkinson,$^{41}$
and J.~Zuntz,$^{47}$
\begin{center} (DES Collaboration) \end{center}
}
}

\date{Accepted XXX. Received YYY; in original form ZZZ}

\pubyear{2020}

\begin{document}
\label{firstpage}
\pagerange{\pageref{firstpage}--\pageref{lastpage}}
\maketitle

\begin{abstract}
For ground-based optical imaging with current CCD technology, the
Poisson fluctuations in source and sky background photon arrivals
dominate the noise budget and are readily estimated.  Another
component of noise, however, is the signal from the undetected
population of stars and galaxies.  Using injection of artifical
galaxies into images, we demonstrate that the measured variance of galaxy
moments (used for weak gravitational lensing measurements) in Dark
Energy Survey (DES) images is significantly in excess of the Poisson
predictions, by up to 30\%, and that the background sky levels are
overestimated by current software.  By cross-correlating distinct
images of ``empty'' sky regions, we establish that there is a
significant image noise contribution from undetected static sources (US),
which on average are mildly resolved at DES resolution.  Treating these
US as a stationary noise source, we compute a
correction to the moment covariance matrix expected from Poisson noise.  The
corrected covariance matrix matches the moment variances
measured on the injected DES images to within 5\%.  Thus we have an empirical method to 
statistically account for US in weak lensing measurements, rather than requiring
extremely deep sky simulations.  We also find that local sky determinations can remove the 
bias in flux measurements, at a small penalty in additional, but quantifiable, noise.
\end{abstract}




\section{Introduction}
\label{sec:intro}

Optical images from modern astronomical surveys are subject to noise coming both from the detector and from shot noise of arriving source and background photons. By taking calibration data, we can accurately measure the detector read noise, and calculate the detector gain to yield an accurate Poisson noise estimate.  Nuisance signals such as cosmic rays or satellite trails can be identified with streak finders and excised from the data. Together these standard techniques yield an estimate of the total noise in the image.  Many astrophysical investigations require very accurate estimation of the image noise and background levels in order to obtain unbiased inferences. We will focus on weak graviational lensing (WL) measurements of the shapes of galaxies, but other astrophysical investigations, e.g. searches for flux variability, depend heavily on accurate knowledge of the \emph{uncertainties} in source measurements.

There is, however, an additional source of noise in background-limited images that is typically ignored: the contribution from undetected background (or foreground) sources (US), which will add noise above the Poisson expectation for the mean background flux.  Some studies have examined the effect of US on specific methods of WL shear measurement. For example, \citet{2017MNRAS.468.3295H} find that calibration simulations must include undetected background galaxies with $m_{F606W}$$\sim29$ to ensure calibrated multiplicative biases $<1\times10^{-4}$ for the "KSB" estimator of WL shear \citep{1995ApJ...449..460K}.  For the IM3SHAPE estimator \citep{2013MNRAS.434.1604Z}, \citet{2018MNRAS.475.4524S} find that the contribution to the bias from undetected background galaxies is well below statistical uncertainties for the year 1 analysis of the Dark Energy Survey (DES; \citealp{2018ApJS..235...33D}). Most recently, \citet{2019A&A...627A..59E} show that undetected background galaxies with magnitude down to $\sim$28 must be included to calibrate the shear measurement bias for three methods: KSB, Source Extractor+PSFEx \citep{bertinShapes}, and MomentsML \citep{2019A&A...621A..36T}.

Here we propose a different way to characterize the contribution of US in the context of the Bayesian Fourier Domain (BFD) shear measurement method \citep{2014MNRAS.438.1880B,2016MNRAS.459.4467B}. The BFD method is a rigorous Bayesian shape measurement algorithm that is unbiased and does not require simulations for calibrations. Unlike other shear measurement algorithms, BFD does not produce point estimates of shape, but instead estimates the shear given each source's image data by comparing to an unsheared prior population of noiseless "template" galaxies (typically drawn from the deep survey within a typical WL photometric survey).  BFD compresses the information on each galaxy to its image moments of order 0, 1, and 2.  These moments are measured in Fourier domain, after correction for the point-spread function (PSF), so that the mean result is independent of image seeing.  These BFD observables are chosen to be linear in the pixel values, enabling accurate propagation of the image noise into a known multivariate Gaussian distribution for the vector of BFD moments.

In this work, we examine the noise distributions for BFD moment measurements in real DES data by adding artificial galaxies to the real images, then measuring their moments.  The DES image-injection process is known as ``Balrog'' \citep{2016MNRAS.457..786S, balrog2}. 
We first show that there is unaccounted-for noise by looking at the BFD moment distribution of injected galaxies. We then measure the statistics of US in DES images by cross-correlating distinct exposures of  ``empty'' sky, i.e. regions where no sources are above the detection threshold.
In this cross-correlation between exposures,  all temporally stochastic noise sources, namely read noise and shot noise, will average to zero.
This technique is similar in spirit to the surface brightness fluctuation distance measurement technique \citep{1988AJ.....96..807T} and to estimations of the contribution of high-redshift sources to the luminosity function \citep{2005Natur.438...45K,2012ApJ...753...63K,2013MNRAS.432.3474C}.  Examination of the injected moments also reveals a bias in the background (sky) level estimation in the current DES pipeline.

Second we show that the US noise can be treated as a quasi-stationary noise source that simply adds to the covariance matrix of BFD moments computed from shot noise. The distribution of moment noise measured from injection of artificial images into the DES data is shown to be in good agreement with the multivariate Gaussian distribution described by this augmented covariance matrix.
This treatment of the US noise is ideal because 1) it uses the data themselves to measure the contribution rather than relying on simulations of unknown fidelity and 2) it includes the US contribution as a source of noise organically within the Bayesian calculation, rather than trying to calibrate a bias term after the fact.

In \S 2, we present a summary of the DES data, the Balrog image-injection program, and the BFD shear measurement algorithm. We then describe our characterization of image noise using the BFD moments. In \S 3, we first show that there is excess noise and bias in the BFD moment distribution using injection tests. We also examine the behavior of the sky-level bias, and find it substantially reduced by a per-object local sky estimation. We then cross-correlate images of empty sky, revealing the buried US signal. Using these same empty regions, we define a cross-covariance matrix for BFD moments that can be added to the shot noise matrix, improving our estimate of the variance in BFD moment distributions.   In \S4 we provide a short description of the properties of the US population and discuss some of the assumptions made in this work. In \S5 we summarize this work and describe the path forward for the BFD shear measurement algorithm.

\section{Data \& Methods}
\label{sec:data}

In this section we describe the data and methods used to characterize the noise properties of DES images.

\subsection{DES Y3 Data}
\label{sec:desdata} 

DES is a 6-year program to image $\sim$5000 deg$^2$ of the sky using the Dark Energy Camera installed on the Blanco telescope. The survey is conducted in the $g$, $r$, $i$, $z$, and $Y$ bands aiming for a nominal depth of $\sim$24th magnitude. For this work, we use the data obtained through year 3 of the program (Y3 data). These data cover the entire DES footprint with typically $\sim5$ exposures (of a final total of $\sim$10 exposures) per filter per region of the sky. The rough 10$\sigma$ point source limiting magnitudes are 24.2, 24.0, 23.5, and 22.8 in the $g, r,  i$ , and $z$ bands respectively. We exclude the $Y$ band from this analysis since it is much shallower than the $griz$ data.

DES images are processed in several steps as described in \citet{2018PASP..130g4501M}. First the images undergo several pre-processing steps including cross-talk, overscan, bias, and flat field corrections. Next the pipeline applies astrometric solutions, performs sky background subtraction, identifies and masks cosmic rays and satellite streaks, and finally detects objects from a $gri$ coadd image via Source Extractor \citep{1996A&AS..117..393B}. The data are output into a Multi-epoch Data Structure (MEDS, \citealp{2016MNRAS.460.2245J}) consisting of postage stamps (typically $32\times32$ pixels, or 8\farcs5  square) and basic data for each detected object. Photometric measurements are produced by multi-epoch, multi-object fitting (MOF) \citep{2018ApJS..235...33D}.

Of most importance to this work is the sky background subtraction routine performed on the single-epoch images. A sky background ``template'' set is derived as the first four principal components of a set of $\approx$1000 images taken in a given filter and observing season.  For an individual exposure, a weighted sum of these four templates is constructed to best match the observed background (after reducing the image and templates to the medians of 128$\times$128-pixel regions).  This weighted sum is then subtracted from the image, and
the inverse number of detected sky photons in the background model (plus a contribution from read noise) is saved as the weight map (inverse variance) of the image.  This ``PCA'' background is hundreds to thousands of photoelectrons per pixel, depending on the filter, lunar phase, etc.

A subsequent background estimation algorithm is applied to each exposure during 
the cataloging process.  This step is performed by the \textsc{SExtractor} sky-estimation algorithm in its \texttt{GLOBAL} mode, whereby medians of regions of size $\texttt{BACK\_SIZE}=256$-pixel square regions are arrayed, smoothed with a $3\times3$ median filter, and interpolated back to single-pixel resolution [see \citet{2018PASP..130g4501M} for full details on \textsc{SExtractor} parameter settings].  The \textsc{SExtractor} sky estimation is needed to account for scattered light from the brightest stars, Galactic dust, and other artifacts that are specific to individual pointings and not captured by the PCA.  These corrections are typically $O(10)$ photoelectrons per pixel or less, i.e. a small perturbation to the PCA sky.
For each exposure/object combination, a postage stamp image is saved to the MEDS file, which already has the
PCA sky and the \textsc{SExtractor} sky estimate subtracted.  The standard procedure for analyzing images is to assume that the MEDS stamp has zero background.

\subsection{Balrog}
\label{sec:balrog}

The Balrog pipeline aims to assess detection efficiency, selection biases, and other biases by injecting fake galaxies with known input parameters into real DES images and running the object detection and photometry as for the real data \citep{2016MNRAS.457..786S}. The current Balrog analysis for DES Y3 injects parametric galaxy models, with the population of injection galaxies drawn from MOF fits to galaxies found in the DES deep fields, as detailed in \citet{balrog2} and \citet{deepfields}.  These deep-field images are produced by summing the many exposures taken by DES in each of the 10 supernova-search fields and in the COSMOS field.  The deep fields are $\sim$1 mag deeper than the Y3 DES coadds. The injections are done on the single-epoch images, which are then run through the processing steps for coaddition, object detection, MEDS making, and photometry outputs such as MOF.  DES coadd creation, and the Balrog processing, are executed in units of 0.5-square-degree patches of sky known as ``tiles.''
In this work we make use of 48 tiles for which the full Balrog injection and reanalysis have been completed.
We label these as the ``Balrog-injection'' tiles.  An average of $\sim3000$ Balrog injections are detected in each tile, though only a subset of them will pass the isolation and $S/N$ cuts that we impose for measurements in this paper.  

We also run a variant of the Balrog pipeline to produce MEDS files containing nominally empty patches of sky. This is done by running the Balrog pipeline but skipping the step where the galaxies are actually added to the images. Postage stamps of these ``ghost'' Balrog injections thus do not contain any central injected galaxy,  but may include real galaxies that were located nearby.  We discard any ghost stamps which are
located close enough to a detected real object that its detected isophotes impinge on the stamp, leaving us with a MEDS file of apparently source-free but otherwise random regions of sky. We label these DES tiles as "Balrog-variant" tiles, of which we have 39 for analysis in this work.
\subsection{BFD}
\label{sec:bfd} 

The BFD shear measurement algorithm is a rigorous Bayesian computation of shear given the data for an ensemble of galaxies. It does not require simulations for calibrating biases. The method compresses pixel-level data to 7 moments computed in Fourier space. A template population of galaxies measured in low-noise imaging serves as prior knowledge on the galaxy population in this moment space. The heart of BFD is to integrate each target galaxy's measured likelihood of moments against a prior embodied by a sheared version of the template population, to produce a posterior probability for the shear given the observed moments. The individual galaxies' shear posteriors are multiplied to obtain the probability of shear given the full galaxy sample.

The 7 BFD moments are defined as:
\begin{equation}
    M \equiv \begin{pmatrix}
    M_F\\
    M_X\\
    M_Y\\
    M_R\\
    M_1\\
    M_2\\
    M_C\\
    \end{pmatrix}
    = \int d^2k \frac{\tilde{I}(k)}{\tilde{T}(k)} W(k^2) F;
    F = \begin{pmatrix}
    1\\
    k_x\\
    k_y\\
    k_x^2+k_y^2\\
    k_x^2-k_y^2\\
    2k_xk_y\\
   (k_x^2+k_y^2)^2\\
    \end{pmatrix}
\label{eq:moments}
\end{equation}

\noindent where $\tilde{I}(k)$ is the Fourier transform of the galaxy postage stamp, $\tilde{T}(k)$ is the Fourier transform of the PSF, and $W(k^2)$ is a weight function designed to prevent the integral from going to infinity where the PSF goes to zero. $M_F$ is the zeroth-order flux moment; $M_X$ and $M_Y$ are the first-order centroid moments; $M_R$, $M_1$, and $M_2$ are the second-order shape/size moments, and $M_C$ is a fourth order moment approximating concentration. In practice we use fast Fourier transforms (FFTs) to obtain the Fourier transform of the postage stamp and the integral becomes a sum over k-space. When measuring the moments for a galaxy, we centroid on the galaxy by zeroing the first order $M_X$ and $M_Y$ moments and then measure the five other moments.  In the standard processing, the level of sky in the MEDS postage stamp is assumed to be zero, i.e. \textsc{SExtractor} sky estimate is correct---which affects only the $k=0$ element of $\tilde I.$  We will investigate an alternative sky subtraction in Section~\ref{sec:balnoisetests}.

One major assumption of BFD is that the pixel noise in the image is stationary and the probability distribution of the observed moments can be described as a multi-dimensional Gaussian about the true moments, with covariance matrix elements for the $i$th and $j$th moment that can be computed from the power spectrum of the noise $P_n(k)$:
 
 \begin{equation}
   \mathbf{Cov}_M[i,j] = \int d^2k P_n(k)
   \left|\frac{W(k^2)}{\tilde{T}(k)}\right|^2 F_i(k) F_j(k)
 \label{eq:covshot}
 \end{equation}
 
 \noindent In the case of sky background and detector read noise, which should both have white-noise spectra, these conditions are upheld. In the presence of significant shot noise from the source itself, the assumption is not valid. For weak lensing, the majority of our signal comes from faint galaxies where the source shot noise should be  insignificant. 

In \citet{2016MNRAS.459.4467B}, it was shown that the BFD method could produce a nearly unbiased result on postage stamp simulations. There was a remaining multiplicative bias of $\sim$0.002 that was not explained. In the Appendix, we present updated validation simulations, showing that the method is unbiased within next generation survey goals of $|m| < 0.002$. Briefly, we find that zero-padding the images (i.e. augmenting the image with regions of zero flux before conducting the Fourier transform) produces more accurate measurements of the moments and their derivatives under shear, since zero-padding produces finer sampling (and better interpolation) of the Fourier space image.


\subsection{Noise Tests}
\label{sec:noisetests}

The accuracy of BFD shear estimates has been assessed in image simulations in which the noise is constructed to be stationary and Gaussian.  
These conditions must be verified in real data. The Balrog simulations are perfect for validating these assumptions for BFD in a real data setting, since we can measure the true moments $M_T$ of injected galaxies and compare with their measured moments $M_D$ relative to the noise distribution expected from the known levels of sky background noise and detector read noise.

We conduct two tests in this vein. First we define the quantity $\chi_M$ for a particular moment $M$ with expected standard deviation $\sigma_M$, often called the ``pull'':

\begin{equation}
    \label{eq:chi}
    \chi_M = \frac{M_D - M_T}{\sigma_M}
\end{equation}

\noindent If our noise estimate $\sigma_M$ is correct, then the distribution of $\chi_M$ should follow that of a unit normal ($\mu$ = 0.0, $\sigma$ = 1.0). If we see that the $\sigma$ of our distributions is larger than 1.0, then there is some extra noise component unaccounted for in our data. If it is smaller than 1.0, then we might suspect that the noise level has been overestimated.

Alternatively, we can examine the distribution of $\chi_M^2$, which should follow a $\chi^2$ distribution with one degree of freedom (also yielding $\langle \chi^2_M\rangle=1$). This alternative test can be generalized to $N$ galaxy moments: 

\begin{equation}
  \label{eq:chiN}
  \chi^2_M =  \left(\vec{M}_D - \vec{M}_T\right)^T {\mathbf{Cov}_M}^{-1} \left(\vec{M}_D - \vec{M}_T\right)
\end{equation}

\noindent where $\vec{M}_D$ and $\vec{M}_T$ are vectors containing $N$ BFD moments. The generalized $\chi_M^2$ distribution should follow a $\chi^2$ distribution with $N$ degrees of freedom and yield $\langle \chi^2_M\rangle = N$.

\section{Noise Properties of DES Images}
\label{sec:noiseprops}

In the following sections we examine the noise properties of DES images. We start by performing the noise tests described in \S \ref{sec:noisetests} on the Balrog tiles, finding excess noise. We then look at the cross-correlation of blank sky regions to examine whether this excess noise is due to US. Finally, we characterize the US noise contribution and suggest a way to mitigate its effect for the BFD method.

\subsection{Balrog Tests}
\label{sec:balnoisetests}

\begin{figure}
    \includegraphics[width=\columnwidth]{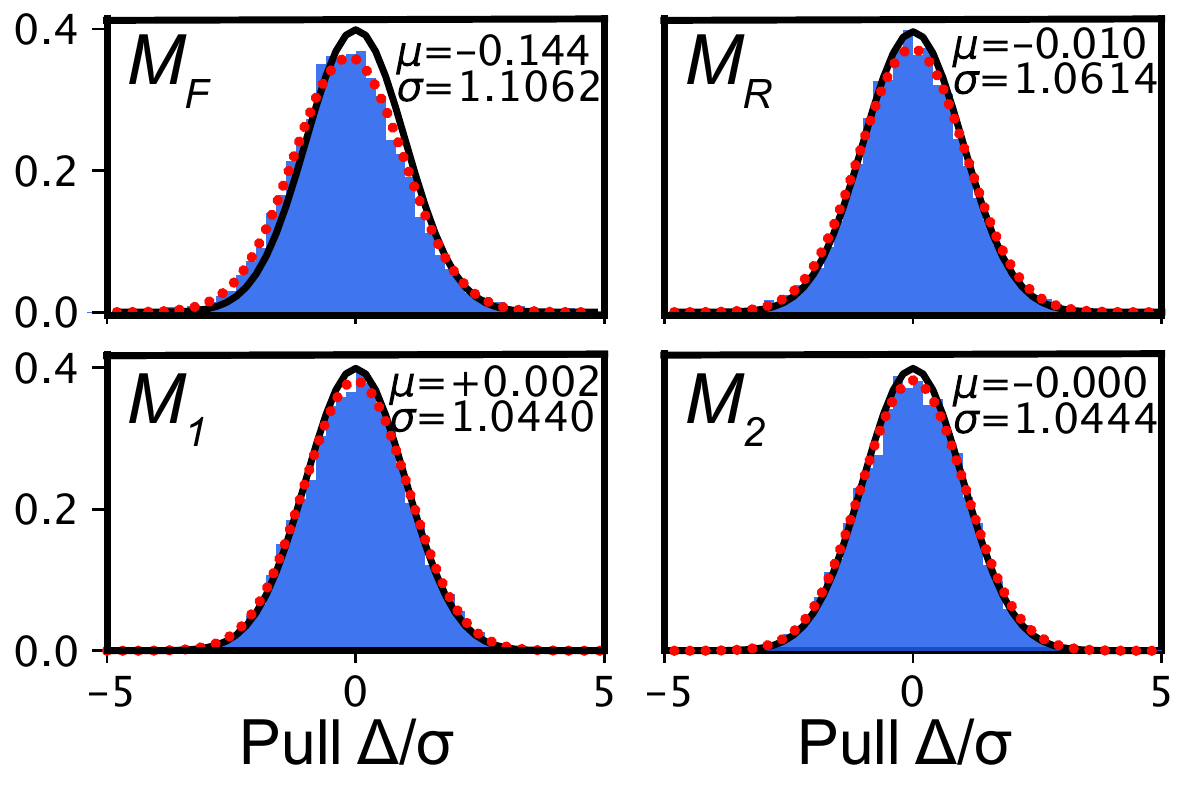}
    \caption{Histograms of $\chi_M$ for four BFD moments a) $M_F$ b) $M_R$ c) $M_1$ and d) $M_2$ for one DES Balrog-injection tile in the $i$~band. The $\mu$ and $\sigma$ values for a Gaussian fit to each moment's data are given and shown by the red dotted curve. The black curve is the unit Gaussian. The flux moment shows the largest deviation from a unit Gaussian. In each histogram there are $\sim$7000 data points, thus the formal 1-$\sigma$ uncertainties on the mean and standard deviation are $\sim$0.01 and $\sim$0.008 respectively. }
    \label{fig:balroghisttest}
\end{figure}

\begin{figure}
    \includegraphics[width=\columnwidth]{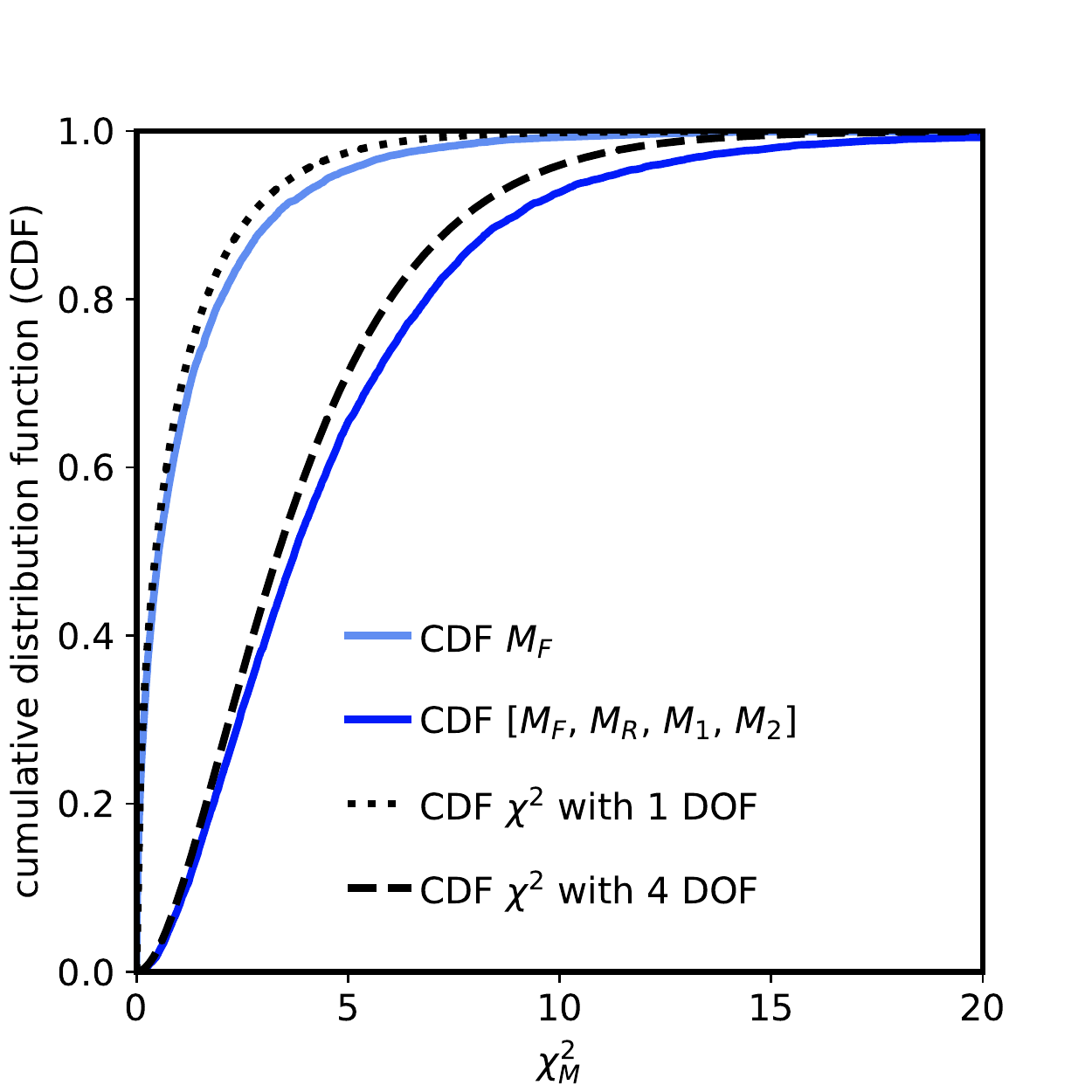}
    \caption{Cumulative distribution function (CDF) of $\chi_M^2$ for $M_F$ (light blue) and $\vec{M}$ = [$M_F$, $M_R$, $M_1$, $M_2$] (dark blue) with $\langle \chi^2_M\rangle = 1.24$ and 4.64 respectively. For comparison, we plot the cumulative distribution functions of a $\chi^2$ distribution with $N = 1$ degree of freedom (dotted black line) and with $N = 4$ degrees of freedom (dashed black line). It is clear that the distribution of $\chi_M^2$ is not consistent with the hypothesis that the noise is purely due to shot noise from the sky background and detector read noise. }
    \label{fig:balrogchi2test}
\end{figure}

To characterize the noise in DES images, we perform the tests described in \S \ref{sec:noisetests} for 48 Balrog-injection tiles randomly selected from the full DES footprint. In Figures \ref{fig:balroghisttest} and \ref{fig:balrogchi2test}, we show the results of the two noise tests for one DES Balrog-injection tile (DES0332-3206) in the $i$~band. For these tests we select sources with S/N in $M_F$ between 3--20 and having no detected neighbors closer than 5\arcsec. These requirements leave us with$\sim$7000 separate images of the injected galaxies, which counts images of the same galaxy injected on different exposures as distinct.
Across all 48 Balrog-injection tiles, there are $\sim$300,000 injections meeting those requirements. In Figure \ref{fig:balroghisttest}, we show the histograms of $\chi_M$ for four BFD moments, the zeroth-order flux moment ($M_F$) and the three second-order shape moments ($M_R$, $M_1$, $M_2$) along with the Gaussian fit to the data (red dotted line) and the unit normal distribution (black).  In Figure \ref{fig:balrogchi2test} we show the cumulative distributions of  $\chi_M^2$  for $M_F$ and for all four moments combined compared to the $\chi^2$ cumulative distribution with $N = 1$ and $N = 4$ degrees of freedom. Tables \ref{tb:chitesttable} and \ref{tb:chi2testtable} give the numerical values of the fits to $p(\chi_M)$ and the $\langle \chi^2_M \rangle$ values, respectively, for all 48 Balrog-injection tiles combined.

The widths of the distributions of $\chi_M$ are generically larger than $\sigma = 1.0$, which we observe across all DES Balrog-injection tiles and $griz$ bands, indicating 6--30\% underestimates of the pixel variance. The width is largest for $M_F$, the flux moment, but is also significant in the 2nd-order shape moments, particular $M_R$. In addition, we see that there is a small but significant sky oversubtraction of $\approx0.13\sigma,$ resulting in a negative $\mu$ offset in $M_F$, which is the only moment sensitive to a global sky offset because it is the only moment with non-zero weight on the DC term of the Fourier space image ($k=0$).

The same trends are reflected in the $\chi_M^2$ test, where we see that the data are not consistent with $\chi^2$ distribution of $N = 1$ or $N = 4$ degrees of freedom. This result suggests that there is additional noise contributing to our data which is not included in our measurement of the pixel noise due to the sky background and detector read noise.

\begin{figure}
    \includegraphics[width=\columnwidth]{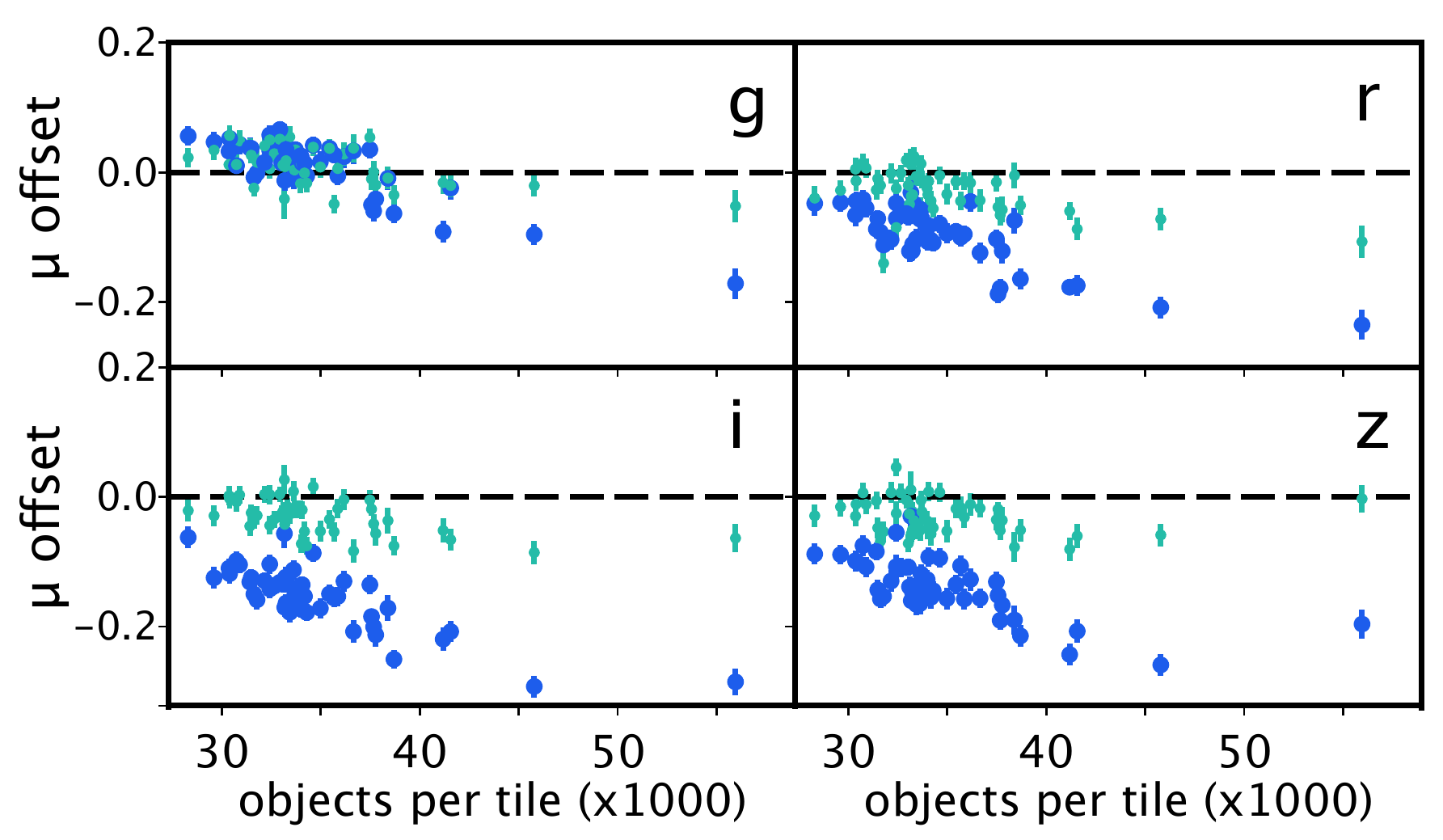}
    \caption{ $\mu$ offset (as calculated in Figure \ref{fig:balroghisttest}) as a function of object density per DES tile for 48 Balrog-injection tiles (blue). The four panels show data in $griz$ bands. While $g-$band is relatively unaffected, we demonstrate that the sky has been oversubtracted in the $riz$ bands, which tends to increase in areas of higher object density. The green dots show the $\mu$ offset after performing a local sky subtraction, which mitigates the flux bias.}
    \label{fig:skysub}
\end{figure}

Exploring further the sky oversubtraction, we show in Figure \ref{fig:skysub} the $\mu$ offset from the noise test in each band as a function of object density per DES tile (object counts with $m_i$ < 23.5) for the set of 48 Balrog-injection tiles. It is apparent that the $g-$band is relatively unaffected, but moving to redder bands, we find larger offsets, which increase with object density. The mean $\mu$ offsets from all 48 Balrog-injection tiles are $\mu_g = 0.023$, $\mu_r = -0.08$, $\mu_i = -0.135$, and $\mu_z = -0.123$.
We suspect that this oversubtraction may be due to: 1) residual light from large, bright galaxies in the frame (similar to what was noted in \citealp{2011AJ....142...31B}); 2) scattered-light halos from stars;  and/or 3) errors in the \textsc{SExtractor} algorithm's treatment of the US component of background. 

To mitigate this bias for BFD, we perform a local sky background subtraction by measuring the mean local background in a 2-pixel-wide frame around the postage stamp. We then convert this background to an offset in the flux moment $M_F$. We also compute the contribution to the variance in $M_F$ by computing the uncertainty due to subtracting this local sky value from each pixel in the postage stamp. The $\mu$ offset for each Balrog-injection tile after performing subtraction is shown in Figure \ref{fig:skysub} (green), and the mean $\mu$ offsets from all 48 tiles are much closer to zero: $\mu_g = 0.019$, $\mu_r = -0.023$, $\mu_i = -0.022$, and $\mu_z = -0.025$. Removing the local background adds noise to the flux moment $M_F$, so we correct the shot/read noise variance for $M_F$ by adding the variance from the sky measurement uncertainty.   Thus the local sky subtraction appears to reduce the mean sky error from 10--20\% of the sky noise for the \textsc{SExtractor} estimate to $\approx 2\%$ of sky noise, which is $\lesssim 1$ photoelectron.   The nature of the residual biases are not yet understood.

\subsection{Empty Sky Cross Power Spectra}
\label{sec:emptysky}

To assess the US noise contributing to our data, we use the 39 Balrog-variant tiles described in \S \ref{sec:balrog} to isolate postage stamps with empty sky. Within each tile, we find the postage stamps belonging to Balrog galaxies (that were not actually injected) and cut down the postage stamp to 32x32 pixels (the minimum postage stamp size for DES galaxies). Second, we require that no neighbors (detected, real galaxies) have \textsc{SExtractor} isophotes extending within the boundaries of the postage stamp, as indicated by the \textsc{SExtractor} segmentation maps.

For each  postage stamp of empty sky, we measure the auto-power spectra for each single-epoch image and the cross-power spectra of all combinations of the single-epoch images in the same filter. For $N>1$ images, this gives $\frac{N!}{2!(N-2)!}$ possible combinations to produce cross power spectra. Shot noise and read noise should average to zero in the cross-power spectra. If there are US below the detection threshold, they will contribute coherently to all single-epoch images and yield positive cross-power spectra.

\begin{figure}
    \includegraphics[width=\columnwidth]{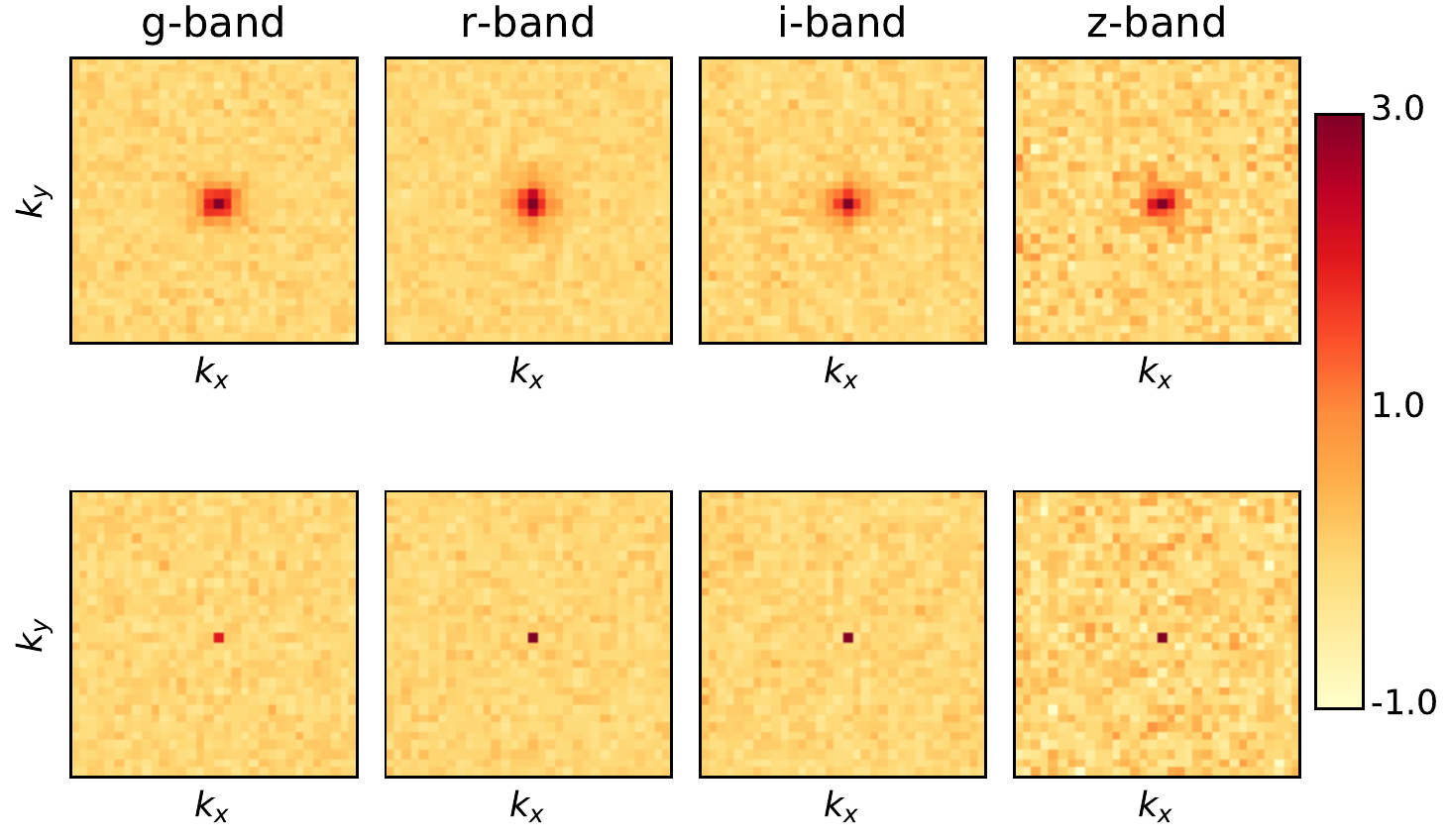}
    \caption{2D average cross-power spectra of empty sky regions in $griz$ bands for one DES Balrog-variant tile normalized to the 99th percentile power in each band (bottom row is normalized to the same value as the top row). Each panel has $k=0$ at the center, and the Nyquist frequency (0.5 cycles per pixel) at the borders. Top: Cross-power spectra are taken for all single-epoch images of one region, thus sky background and detector noise should cancel, isolating any contribution from US. Bottom: Cross-power spectra are taken for single-epoch images of different regions, thus sky background and detector noise, as well as US should cancel.  
Only background estimation errors, at $k=0,$ remain non-zero.}
    \label{fig:powerspectra}
\end{figure}

To look for the US signal, we average the cross-power spectra over the entire Balrog-variant tile. The 2D average cross-power spectra for each band of one tile are shown in the top row of Figure \ref{fig:powerspectra}. There is clearly a US signal present in all bands. To ensure that this is not an artifact of the instrument or detector, we also measure the cross-power spectra of distinct patches of blank sky (where the US signal as well as shot noise should average to zero). These 2D average cross-power spectra are shown in the bottom row of Figure \ref{fig:powerspectra}, and it is evident that there is no signal in this case.  The only significantly non-zero cross-power is the DC term ($k=0$), which arises due to the systematic tendency to over-subtract the background.

\begin{figure}
    \includegraphics[width=\columnwidth]{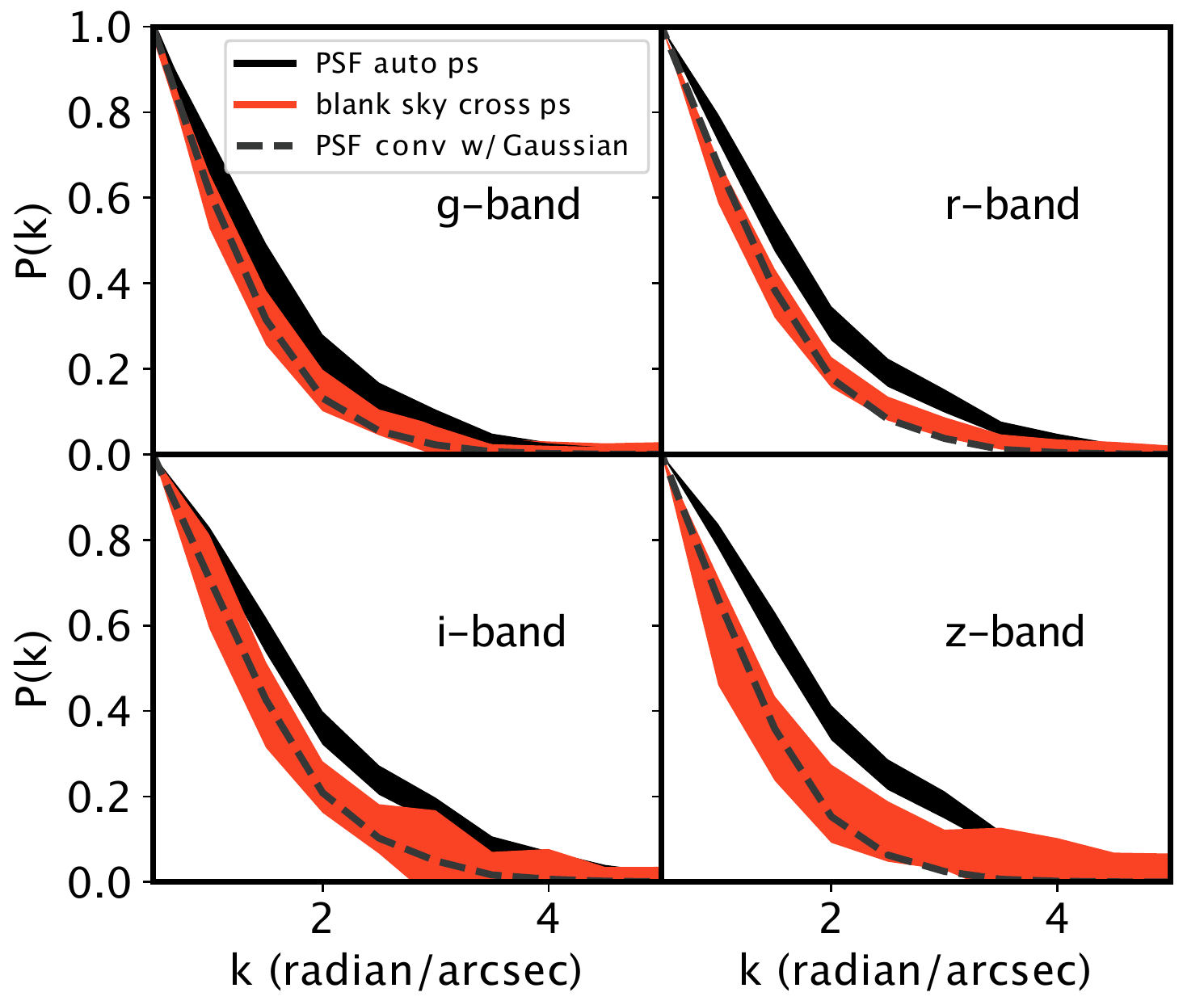}
    \caption{Profiles of cross-power spectra of empty patches for the $griz$ bands averaged over 39 Balrog-variant tiles. The cross-power spectra are normalized to the power at $k = 0.05$. The orange shading shows the mean profile from all 39 tiles, with width representing the standard deviation. The black shading shows the mean profile of the average PSF auto-power spectrum for each tile, with width representing the standard deviation. The grey dashed line shows the mean PSF profile convolved with a Gaussian of given $\sigma$, to estimate the rough size of the average US population. We find $\sigma$ $\approx$ 0.25, 0.25, 0.25, and 0.32 arcseconds for  $g$, $r$, $i$, and $z$ respectively.} 
    \label{fig:powerspecprofiles}
\end{figure}

The azimuthally averaged profiles of the cross-power spectra from empty regions of 39 Balrog-variant DES tiles are shown in Figure \ref{fig:powerspecprofiles}, along with the average auto-power spectra of the PSF of each tile (The shading represents the standard deviation of profile shapes from the 39 different tiles). The profiles are all normalized at $k=0.5,$ since the $k=0$ term has extra noise due to sky background subtraction errors. We find that the averaged US population is resolved in all bands. To estimate the average size of US in each band, we convolve the PSF with Gaussians of varying $\sigma$ and compare the resulting power spectrum with the average cross-power spectra of US (black dashed line). We find that $\sigma = 0.25\arcsec$ for the $g$, $r$, and $i$ bands and $\sigma = 0.32\arcsec$ for the $z$ band yield rough agreement with the US profile. Within the uncertainties of this measurement, the cross-power spectra are consistent with arising from a population of sources with a slightly resolved profile (intrinsic FWHM$\approx0.7\arcsec,$ with the typical PSF of DES imaging having FWHM$\approx0.9\arcsec$ in $riz$ bands).
It is not entirely surprising that the sizes should differ across bands, as the populations of US probed in each band will be slightly different.  

\subsection{Computing Cross-Covariance}
\label{sec:crosscov}

We can treat the background source population as a roughly stationary noise source.  While this should be strictly true for our randomly-placed Balrog-injected galaxies, it will not be precisely true for real galaxies, for reasons we discuss in Section \ref{sec:discussion}. We will proceed, however, to compute a new covariance matrix for the US noise which we will add to our nominal shot/read noise covariance matrix.

We could in principal use the cross-power spectrum computed in \S \ref{sec:emptysky} to compute the US covariance of the moments, as described in equation \ref{eq:covshot} (see also equation 9 of \citealp{2016MNRAS.459.4467B}). However, this cross-power spectrum does not include corrections for the PSF and small pixel shifts for each individual image. These imperfections smear out the signal.

To compute the cross-covariance matrix more accurately we compute the BFD moments for each image of empty sky. Using the BFD software enables us to properly account for the differing PSFs and world coordinate systems (i.e. registration) of each image. We then compute the empirical cross-covariance matrix of BFD moments, $\mathbf{Cov}_X$.

\begin{equation}
\label{eq:crosscoveq}
   \mathbf{Cov}_{X[i,j]} = \frac{\sum_{\alpha,\beta, \alpha \neq \beta} M_{i\alpha}M_{j\beta} }{\sum_{\alpha,\beta, \alpha \neq \beta} 1} 
\end{equation}

\noindent where ($i$, $j$) are indexes of the moment element of the covariance matrix, and $\alpha$ and $\beta$ index the exposures of this sky patch. The summation is performed over all combinations of images where $\alpha$ $\neq$ $\beta$. We combine data from all 39 Balrog-variant tiles to compute a global cross-covariance matrix for the DES data from empty sky regions. 

Finally, we test how using the new cross-covariance matrix affects the noise tests described in \S \ref{sec:noisetests}. Analogous to $\chi_M$ we now define $\chi_{MX}$, which is now calculated using the sum of the shot/read-noise covariance and $\mathbf{Cov}_X$

\begin{equation}
    \label{eq:chinew}
    \chi_{MX} = \frac{M_D - M_T}{\sqrt{\sigma_M^2 + \sigma_X^2}}
\end{equation}

\noindent where $\sigma_M$ is from the shot/read-noise covariance matrix and $\sigma_X$ is from the cross-covariance matrix $\mathbf{Cov}_X$. Generalizing to the $\chi^2$ test using arbitrary moments we have:

\begin{equation}
  \label{eq:chi2new}
  \chi^2_{MX} =  \left(\vec{M}_D - \vec{M}_T\right)^T [{\mathbf{Cov}_M+\mathbf{Cov}_{X}}]^{-1} \left(\vec{M}_D - \vec{M}_T\right)
\end{equation}

\begin{figure}
    \includegraphics[width=\columnwidth]{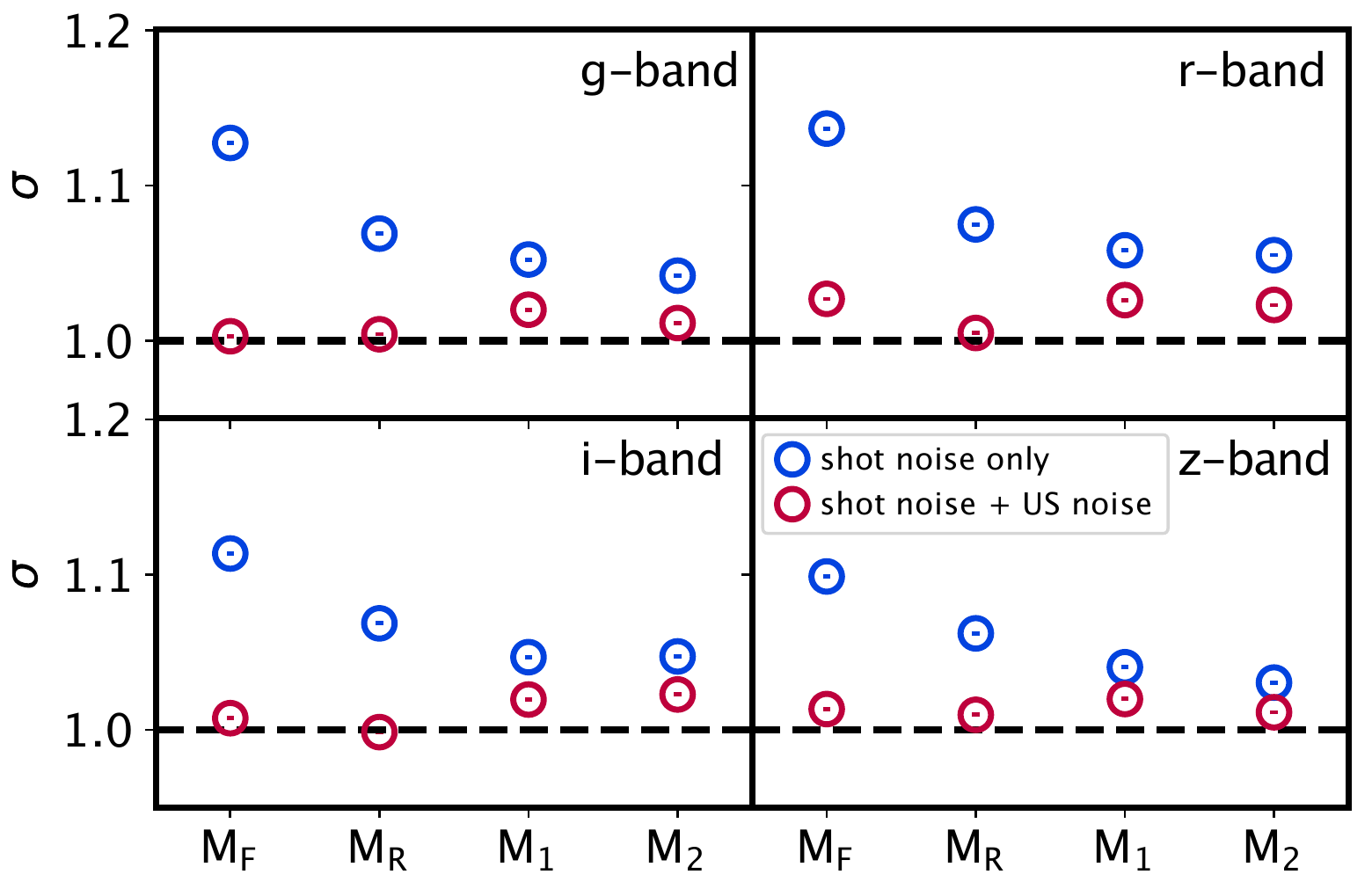}
    \caption{$\sigma$ of $\chi_M$ distribution (blue) and $\chi_{MX}$ distribution (red) for each BFD moment from all 48 Balrog-injection tiles. Including the cross-covariance matrix brings the $\sigma$ of the distribution closer to the expected value of $1$. The values of $\sigma$ for each band/moment combination are given in Table \ref{tb:chitesttable}.}
    \label{fig:sigmatest}
    \end{figure}

For these tests, we combine moments from all 48 Balrog-injection tiles, and compute the $\mu$ and $\sigma$ of the normalized histogram (as in Figure \ref{fig:balroghisttest}) for each band. We have subtracted the local sky estimate for each postage stamp and included the variance due to that local sky in the flux moment variance term of $\mathbf{Cov}_M$. In Figure \ref{fig:sigmatest}, we show the $\sigma$ value for each band/moment combination using  $\mathbf{Cov}_M$ and using $\mathbf{Cov}_M$ + $\mathbf{Cov}_X$. We find general improvement using our globally defined cross-covariance term, with $\sigma$ lowering from $\sim$1.12 to $\sim$1.01 for the flux moment, and dropping to similar levels for the other moments ($\sigma=1.026$ in the worst case).  We report the $\sigma$ and $\mu$ values for all bands in Table \ref{tb:chitesttable}.

\begin{figure}
    \includegraphics[width=\columnwidth]{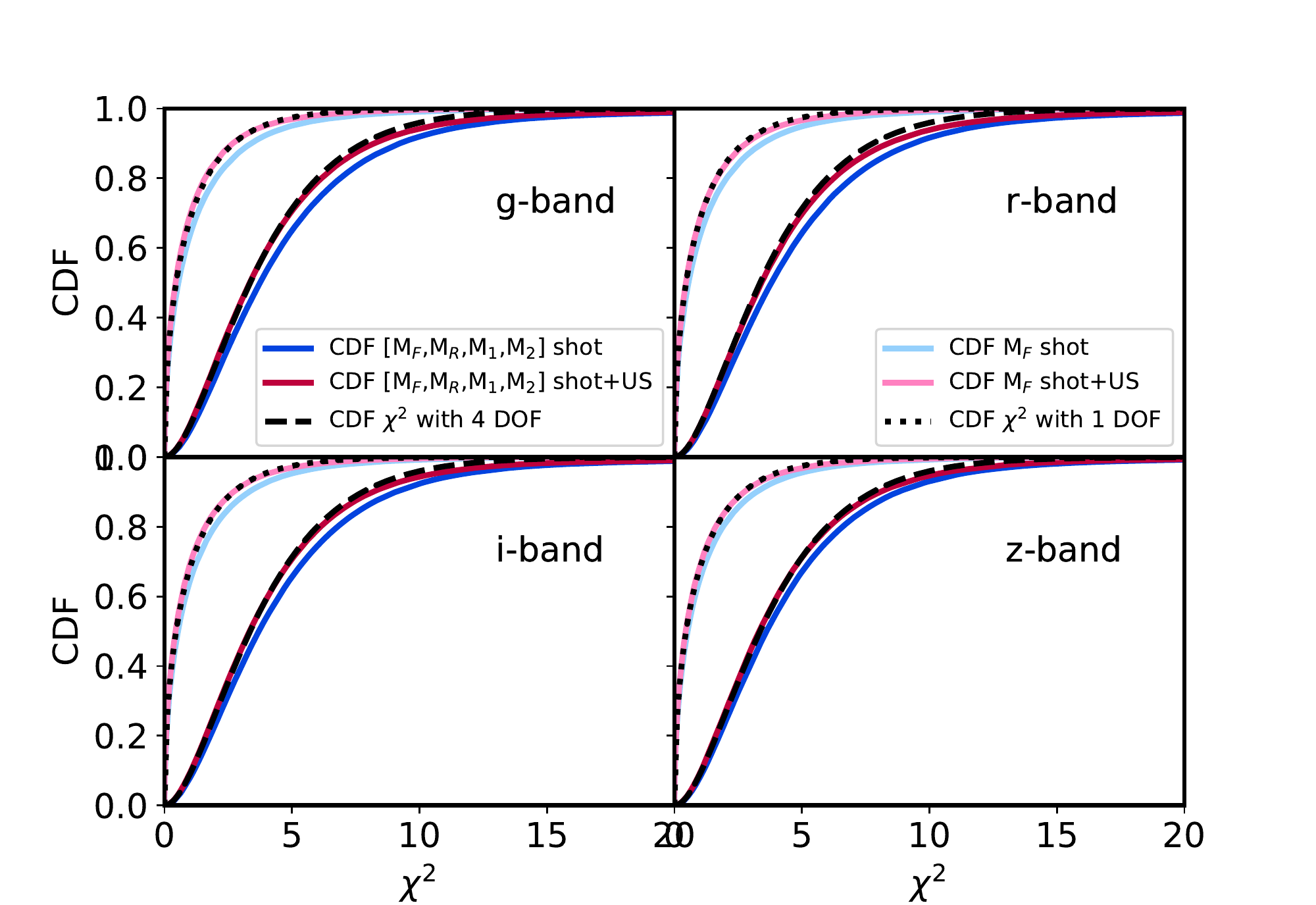}
    \caption{CDF of $\chi_M^2$ (blue/light blue) and $\chi_{MX}^2$ (red/pink) for the single BFD $M_F$ moment and for the four $M_F$, $M_R$, $M_1$, and $M_2$ moments for each band over all 48 Balrog-injection tiles. The $\chi^2$ CDFs with 1 degree of freedom (black dotted line) and 4 degrees of freedom (black dashed line) are plotted for reference. Again, we show that including the cross-covariance term brings the moment distribution closer to expectations. Mean $\chi^2$ values for each distribution are given in Table \ref{tb:chi2testtable}}
    \label{fig:chi2test}
\end{figure}

Furthermore, in Figure \ref{fig:chi2test}, we show the full $\chi^2$ test for $M_F$ only and for all 4 moments of a given band across all 48 Balrog tiles and we report the $\langle \chi^2_M\rangle$ value in Table \ref{tb:chi2testtable}. In all bands, we find that when including the cross-covariance matrix that characterizes the US noise, the CDF of the data more closely approaches the $\chi^2$ distribution with the appropriate number of degrees of freedom. The $\langle \chi^2_M\rangle$ values also reflect the improvement, as they approach the expected value of $1$ for $M_F$ only and $4$ when looking at the vector of 4 moments.

\begin{table*}
	\centering
	\caption{Mean ($\mu$) and standard deviation ($\sigma$) values for noise test 1, with ``shot'' denoting inclusion of background Poisson noise and read noise only, and ``shot$+$US'' case including the measured US noise variance as well. These results include a local sky subtraction as described in Section~\ref{sec:balnoisetests}.  The $\sigma$ values are plotted in Figure~\ref{fig:sigmatest} as well.}
	\label{tb:chitesttable}

\begin{tabular}{lcccc}
\hline
moment & $g$ & $r$ & $i$ & $z$ \\
       & $\mu$\hspace{1.2em}$\sigma$ & $\mu$\hspace{1.2em}$\sigma$ & $\mu$\hspace{1.2em}$\sigma$ & $\mu$\hspace{1.2em}$\sigma$ \\
\hline
$M_F$ & & & & \\
\hspace{0.5em}shot& 0.019\hspace{1em}1.127 & -0.016\hspace{1em}1.136 & -0.022\hspace{1em}1.113 & -0.019\hspace{1em}1.098\\
\hspace{0.5em}shot+US& 0.016\hspace{1em}1.002 & -0.015\hspace{1em}1.026 & -0.020\hspace{1em}1.007 & -0.018\hspace{1em}1.013\\
$M_R$ & & & & \\
\hspace{0.5em}shot& 0.029\hspace{1em}1.068 &  0.017\hspace{1em}1.074 & -0.000\hspace{1em}1.068 & -0.003\hspace{1em}1.062\\
\hspace{0.5em}shot+US& 0.027\hspace{1em}1.004 &  0.016\hspace{1em}1.004 & -0.000\hspace{1em}0.998 & -0.003\hspace{1em}1.009\\
$M_1$ & & & & \\
\hspace{0.5em}shot& 0.001\hspace{1em}1.052 &  0.003\hspace{1em}1.058 & -0.002\hspace{1em}1.046 & -0.003\hspace{1em}1.040\\
\hspace{0.5em}shot+US& 0.001\hspace{1em}1.019 &  0.003\hspace{1em}1.026 & -0.002\hspace{1em}1.019 & -0.003\hspace{1em}1.019\\
$M_2$ & & & & \\
\hspace{0.5em}shot&-0.004\hspace{1em}1.041 & -0.001\hspace{1em}1.055 & -0.001\hspace{1em}1.047 & -0.005\hspace{1em}1.030\\
\hspace{0.5em}shot+US&-0.004\hspace{1em}1.011 & -0.001\hspace{1em}1.023 & -0.001\hspace{1em}1.022 & -0.004\hspace{1em}1.011\\
\hline
\end{tabular}
\end{table*}

\begin{table}
	\centering
	\caption{$\langle \chi^2_M\rangle$ from distributions shown in Figure \ref{fig:chi2test} for each band including shot noise only and shot+US noise, as described in Table \ref{tb:chitesttable}.}
	\label{tb:chi2testtable}
\begin{tabular}{lcccc}
\hline
moment & $g$ & $r$ & $i$ & $z$ \\
       & $\langle \chi^2_M\rangle$ & $\langle \chi^2_M\rangle$ & $\langle \chi^2_M\rangle$ & $\langle \chi^2_M\rangle$ \\
\hline
$M_F$ & & & & \\
\hspace{0.5em}shot & 1.2714&1.2923&1.2406&1.2077\\
\hspace{0.5em}shot+US & 1.0059&1.0548&1.0154&1.0272\\
4$M$ & & & & \\
\hspace{0.5em}shot & 4.6574&4.7239&4.5891&4.5402\\
\hspace{0.5em}shot+US & 4.1515&4.2351&4.1396&4.1811\\
\hline
\end{tabular}
\end{table}

\section{Discussion}
\label{sec:discussion}

The above tests demonstrate that the measured distribution of moment errors is well described by a multivariate Gaussian distribution with covariance matrix $\mathbf{Cov}_M$ that sums the usual contribution from sky/read noise with an additional $\mathbf{Cov}_X$ term that is measured from inter-exposure correlations in ``blank'' sky.  A workable hypothesis is that $\mathbf{Cov}_X$ arises from the signals of undetected sources, i.e. the US noise. In this section we examine the data further to infer the average properties of the US population under this hypothesis as well as discuss the validity of the assumption of stationarity made in this analysis.

\subsection{Properties of Background Galaxies}
\label{sec:bkgdgalprop}

In Section \ref{sec:emptysky}, we found that the US population is mildly resolved with typical sizes of $\sigma$ $\sim$ 0.25\arcsec\ for the $g$, $r$, and $i$ bands, and 0.32\arcsec\ for the $z$ band. Since each band probes different populations of background galaxies (as well as stars), it is not surprising that these values vary. Regardless, our US population consists largely of small galaxies that are slightly larger for redder bands.

As another check of our data, we can also compare the cross-power spectra computed in \S \ref{sec:emptysky} with the predicted power spectra of US from the flux distribution of the DES data, noting that the power spectrum is defined by the flux distribution of sources in the image:

\begin{equation}
\label{eq:pUS}
  P(k) = \int df\, \frac{dn}{df} f^2 |s^2(k,f)|
\end{equation}

\noindent where $dn/df$ is the flux $f$ distribution of sources and $s(k,f)$ is their shape. Assuming that all US are the same small shape, when $k$ is well below the size of the galaxies ($k\rightarrow$ 0), we can define the power spectrum or variance of US:

 \begin{equation}
\label{eq:VUS}
V_{\rm US} = \int df\,\frac{dn}{df} f^2. 
\end{equation}

\begin{figure}
    \includegraphics[width=\columnwidth]{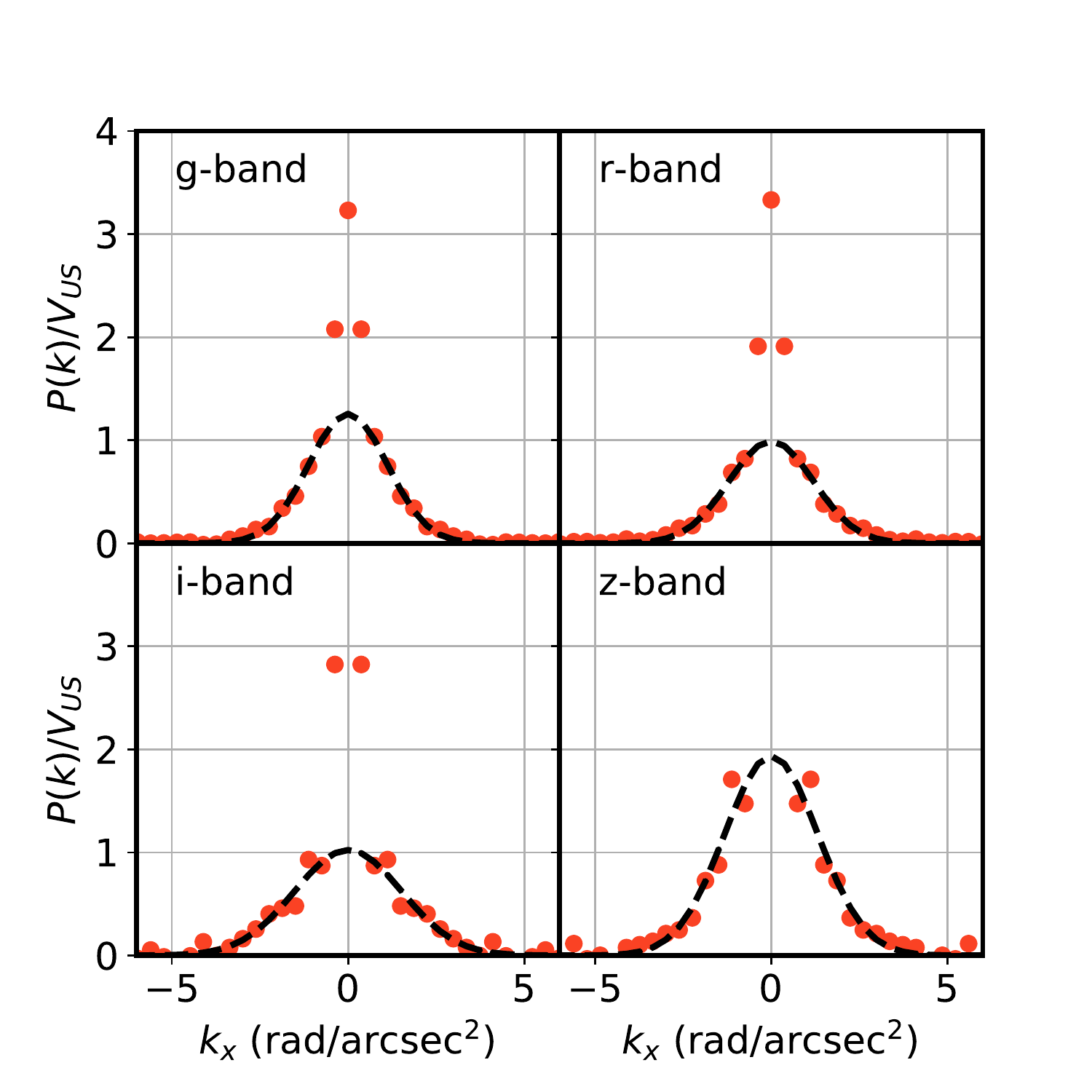}
    \caption{Cut along $k_y=0$ slice of cross-power spectra (as shown in Figure \ref{fig:powerspectra}) normalized by $V_{\rm US}$ as described in \S \ref{sec:bkgdgalprop} for all four bands. The black dashed line shows a Gaussian fit, ignoring the inner three values, which are contaminated by the noise from residual sky background subtraction issues. We find that the cross-power spectra from the data approaches the predicted value of $V_{\rm US}$, although for the $z$ band, it is closer to 2$\times V_{\rm US}$. }
    \label{fig:PkUS}
\end{figure}

Thus to predict $V_{\rm US}$, we need to know the flux distribution of US in the data. To do this, we use equation \ref{eq:VUS} to measure the variance of sources with $i$-band magnitude $> 23.0$ from both the wide field $V_{\rm wide}$ and the deep field $V_{\rm deep}$, which extends $\sim$1 mag deeper than the wide field. We then compute $V_{\rm US} = V_{\rm deep} - V_{\rm wide}$. In Figure \ref{fig:PkUS}, we show the $k_y = 0$ slice of the mean cross-power spectra, normalized by the $V_{\rm US}$ in each band. Performing a Gaussian fit to the cross-power spectra (throwing out the central data points, which are contaminated by the sky background oversubtraction), we find that the normalized cross-power spectra are $\sim$ 1 at $k=0$ (although for $z$ band, it is closer to 2). We expect that the ratio should be larger than one, as we do not extrapolate the flux distribution to fainter magnitudes.

\subsection{Assumption of Stationarity}
\label{sec:stationarity}

In the analysis above, we have assumed that the noise due to US is stationary, i.e., the US galaxies are randomly sprinkled on the sky with respect to our detected population, and thus that the US noise covariance matrix $\mathbf{Cov}_X$ can be simply added to the shot-read noise covariance matrix $\mathbf{Cov}_M$. For the Balrog galaxies, which have been simply added to the DES tiles at a regular hexagonal grid of positions, this assumption is fully valid. However, we know that galaxies are clustered, and that US are clustered around brighter, detected galaxies, breaking our previous assumption. For example, \citet{2019A&A...627A..59E} showed the clustering of galaxies of different magnitudes in the HST Ultra Deep Field (UDF), finding significant excess densities within $\sim$ 2\arcsec of objects with mag(F775W) < 24.5, largely dominated by objects about one magnitude fainter than the chosen threshold. 

Thus we must consider two contributing sources to the US noise in our data: 1) pure projection US (at different redshifts) and 2) physically associated US at the same redshift. For projection US, we expect this signal to be stationary and fully captured by our Balrog simulations. There are some caveats that are not accounted for in Balrog, such as extinction by the foreground galaxy and magnification by the foreground galaxy or other matter along the line of sight. The first caveat would act to decrease the US signal. In nearby spiral galaxies, dust, as traced by the IR, generally follows the optical disk \citep{2013A&A...556A..54V}, although optical disk opacities are generally close to 0 by $R_{25}$ \citep{2005AJ....129.1396H}. The second caveat might increase or decrease the US signal. Regardless these are second order effects on the primary US signal, which contributes $\sim$ 30\% extra variance. At our current 1\% level accuracy in noise determination, these are small effects.

For US that are physically associated with the detected galaxies (same redshift) there is an expected excess around the detected sources, breaking stationarity. \citet{2019A&A...627A..59E} show that introducing clustered US galaxies produces a multiplicative bias $\sim 1\times10^{-2}$ for two shear estimation methods, whereas an unclustered US population induces a bias $\sim 4\times10^{-3}$ (with even the latter being larger than the required accuracy for imminent lensing surveys). The BFD method, however, relies on a template population built from deep sky images. If the template population includes clustered US at similar levels as in the wide field population, BFD will naturally account for the signal due to clustered US in the Fourier moment space. Thus the clustered US population will be treated as signal rather than noise, since it is subject to the same gravitational lensing distortion as the primary detected galaxy.  In other words we can consider the associated US photons to be part of the detected galaxy image.  It will be important, however, to design the detection process such that (un)detected sources remain (un)detected after being subject to weak lensing distortions \citep{2019arXiv191102505S}.  Further simulations will be necessary to quantify the effectiveness of this strategy.


One other aspect of stationarity to consider is whether the derived US-noise matrix $\mathbf{Cov}_X$ is constant across the survey footprint in the face of varying survey noise levels and PSF size.  Nominally the answer is yes, since the BFD moments incorporate a PSF correction, and the shot noise itself is cancelled in the measurement of $\mathbf{Cov}_X.$  It is true, however, that in regions of the survey with lower detection thresholds, there are fewer undetected sources so lower $\mathbf{Cov}_X.$  Thus there may need to be some adjustment of $\mathbf{Cov}_X$ for observing conditions unless we enforce a uniform detection threshold.

\subsection{Metacalibration with undetected sources}
The first-year weak-lensing analyses of the DES data make use of the ``metacalibration'' technique \citep{metacal1,metacal2}, with implementation described by \citet{y1wl}.  Forthcoming DES analyses will use the metadetection technique as well.  Does the undetected-source noise affect metacalibration?  In this technique, a linear operator is applied to the observed image such that all objects in the scene undergo a (pre-seeing) shear operation.  This will naturally apply the same shear to undetected sources as it does to the (targeted) detected sources, thus calibrating their effect correctly, albeit with the assumption that the US undergo identical shear to the target source.  In practice this is not true, for the same reasons that US noise is not stationary, but this is a second-order effect.

The image operator applied during metacalibration also shears the shot noise and read noise in the image.  This is undesirable, since lensing does not shear the noise, and is compensated in the metacalibration algorithm by adding new noise to the image in a manner that re-symmetrizes the total noise (at a penalty in total noise amplitude).  The US noise is not white noise and the re-symmetrization operation would not yield an unsheared realization of US noise, hence it is better considered part of the sheared scene than the unsheared noise.  We thus expect that the presence of US noise does not, to first order, alter the calibrations and results derived from metacalibration.  The sky-level biases discussed herein will, however, have some impact on metacalibration and essentially any WL method that is sensitive to the sky level.  Testing on simulations with undetected sources included indicates, so far, no detectable bias induced by their presence \citep{metadetect}.

\section{Conclusions}
\begin{outline}
We have shown that US produce a significant noise contribution to DES images (up to 30\% in variance), which must be characterized properly for accurate and precise determination of galaxy shapes for shear measurement. Treating the US noise as a stationary noise source, we can empirically compute a cross-covariance matrix for the BFD moments from empty sky regions in DES images. This cross-covariance matrix is then added to the shot/read noise covariance matrix to yield a covariance matrix that describes the observed BFD moment noise distribution to 1--5\% accuracy. Rather than requiring simulations to correct for this noise source, we have shown that BFD can treat the noise statistically, using these corrections derived directly from the sky images.  

While the Balrog simulations do not fully capture the complexities of galaxy clustering vs. pure projection, we suspect that our tests with this simplified population capture most of the problems. US clustered near faint galaxies should be present in both our target and template populations, and thus should be considered as part of the BFD signal, rather than noise. Magnification and dust obscuration are percent level effects on top of the much larger effect described in this work. Future work may explore ways to improve the method shown here for surveys with more stringent requirements than DES.

\end{outline}

\section*{Acknowledgements}

We would like to thank Eric Neilsen and David Stark for helpful discussions on cross-correlating sky images and the dust extent in galaxies, respectively.
KDE and GMB acknowledge support from NASA through a Research Support Agreement with grant
12-EUCLID11-0004, and grants AST-1615555 from the US National
Science Foundation, and DE-SC0007901 from the US Department of Energy.

Funding for the DES Projects has been provided by the U.S. Department of Energy, the U.S. National Science Foundation, the Ministry of Science and Education of Spain, 
the Science and Technology Facilities Council of the United Kingdom, the Higher Education Funding Council for England, the National Center for Supercomputing 
Applications at the University of Illinois at Urbana-Champaign, the Kavli Institute of Cosmological Physics at the University of Chicago, 
the Center for Cosmology and Astro-Particle Physics at the Ohio State University,
the Mitchell Institute for Fundamental Physics and Astronomy at Texas A\&M University, Financiadora de Estudos e Projetos, 
Funda{\c c}{\~a}o Carlos Chagas Filho de Amparo {\`a} Pesquisa do Estado do Rio de Janeiro, Conselho Nacional de Desenvolvimento Cient{\'i}fico e Tecnol{\'o}gico and 
the Minist{\'e}rio da Ci{\^e}ncia, Tecnologia e Inova{\c c}{\~a}o, the Deutsche Forschungsgemeinschaft and the Collaborating Institutions in the Dark Energy Survey. 

The Collaborating Institutions are Argonne National Laboratory, the University of California at Santa Cruz, the University of Cambridge, Centro de Investigaciones Energ{\'e}ticas, 
Medioambientales y Tecnol{\'o}gicas-Madrid, the University of Chicago, University College London, the DES-Brazil Consortium, the University of Edinburgh, 
the Eidgen{\"o}ssische Technische Hochschule (ETH) Z{\"u}rich, 
Fermi National Accelerator Laboratory, the University of Illinois at Urbana-Champaign, the Institut de Ci{\`e}ncies de l'Espai (IEEC/CSIC), 
the Institut de F{\'i}sica d'Altes Energies, Lawrence Berkeley National Laboratory, the Ludwig-Maximilians Universit{\"a}t M{\"u}nchen and the associated Excellence Cluster Universe, 
the University of Michigan, the National Optical Astronomy Observatory, the University of Nottingham, The Ohio State University, the University of Pennsylvania, the University of Portsmouth, 
SLAC National Accelerator Laboratory, Stanford University, the University of Sussex, Texas A\&M University, and the OzDES Membership Consortium.

Based in part on observations at Cerro Tololo Inter-American Observatory, National Optical Astronomy Observatory, which is operated by the Association of 
Universities for Research in Astronomy (AURA) under a cooperative agreement with the National Science Foundation.

The DES data management system is supported by the National Science Foundation under Grant Numbers AST-1138766 and AST-1536171.
The DES participants from Spanish institutions are partially supported by MINECO under grants AYA2015-71825, ESP2015-66861, FPA2015-68048, SEV-2016-0588, SEV-2016-0597, and MDM-2015-0509, 
some of which include ERDF funds from the European Union. IFAE is partially funded by the CERCA program of the Generalitat de Catalunya.
Research leading to these results has received funding from the European Research
Council under the European Union's Seventh Framework Program (FP7/2007-2013) including ERC grant agreements 240672, 291329, and 306478.
We  acknowledge support from the Brazilian Instituto Nacional de Ci\^encia
e Tecnologia (INCT) e-Universe (CNPq grant 465376/2014-2).

This manuscript has been authored by Fermi Research Alliance, LLC under Contract No. DE-AC02-07CH11359 with the U.S. Department of Energy, Office of Science, Office of High Energy Physics.





\bibliographystyle{mnras}
\bibliography{biblio}

\begin{thebibliography}{}
\makeatletter
\relax
\def\mn@urlcharsother{\let\do\@makeother \do\$\do\&\do\#\do\^\do\_\do\%\do\~}
\def\mn@doi{\begingroup\mn@urlcharsother \@ifnextchar [ {\mn@doi@}
  {\mn@doi@[]}}
\def\mn@doi@[#1]#2{\def\@tempa{#1}\ifx\@tempa\@empty \href
  {http://dx.doi.org/#2} {doi:#2}\else \href {http://dx.doi.org/#2} {#1}\fi
  \endgroup}
\def\mn@eprint#1#2{\mn@eprint@#1:#2::\@nil}
\def\mn@eprint@arXiv#1{\href {http://arxiv.org/abs/#1} {{\tt arXiv:#1}}}
\def\mn@eprint@dblp#1{\href {http://dblp.uni-trier.de/rec/bibtex/#1.xml}
  {dblp:#1}}
\def\mn@eprint@#1:#2:#3:#4\@nil{\def\@tempa {#1}\def\@tempb {#2}\def\@tempc
  {#3}\ifx \@tempc \@empty \let \@tempc \@tempb \let \@tempb \@tempa \fi \ifx
  \@tempb \@empty \def\@tempb {arXiv}\fi \@ifundefined
  {mn@eprint@\@tempb}{\@tempb:\@tempc}{\expandafter \expandafter \csname
  mn@eprint@\@tempb\endcsname \expandafter{\@tempc}}}

\bibitem[\protect\citeauthoryear{{Bernstein} \& {Armstrong}}{{Bernstein} \&
  {Armstrong}}{2014}]{2014MNRAS.438.1880B}
{Bernstein} G.~M.,  {Armstrong} R.,  2014, \mn@doi [\mnras]
  {10.1093/mnras/stt2326}, \href
  {http://adsabs.harvard.edu/abs/2014MNRAS.438.1880B} {438, 1880}

\bibitem[\protect\citeauthoryear{{Bernstein}, {Armstrong}, {Krawiec}  \&
  {March}}{{Bernstein} et~al.}{2016}]{2016MNRAS.459.4467B}
{Bernstein} G.~M.,  {Armstrong} R.,  {Krawiec} C.,   {March} M.~C.,  2016,
  \mn@doi [\mnras] {10.1093/mnras/stw879}, \href
  {http://adsabs.harvard.edu/abs/2016MNRAS.459.4467B} {459, 4467}

\bibitem[\protect\citeauthoryear{{Bertin}}{{Bertin}}{2011}]{bertinShapes}
{Bertin} E.,  2011, {Automated Morphometry with SExtractor and PSFEx}.
p.~435

\bibitem[\protect\citeauthoryear{{Bertin} \& {Arnouts}}{{Bertin} \&
  {Arnouts}}{1996}]{1996A&AS..117..393B}
{Bertin} E.,  {Arnouts} S.,  1996, \mn@doi [\aaps] {10.1051/aas:1996164}, \href
  {https://ui.adsabs.harvard.edu/abs/1996A&AS..117..393B} {117, 393}

\bibitem[\protect\citeauthoryear{{Blanton}, {Kazin}, {Muna}, {Weaver}  \&
  {Price-Whelan}}{{Blanton} et~al.}{2011}]{2011AJ....142...31B}
{Blanton} M.~R.,  {Kazin} E.,  {Muna} D.,  {Weaver} B.~A.,   {Price-Whelan} A.,
   2011, \mn@doi [\aj] {10.1088/0004-6256/142/1/31}, \href
  {https://ui.adsabs.harvard.edu/abs/2011AJ....142...31B} {142, 31}

\bibitem[\protect\citeauthoryear{{Calvi}, {Pizzella}, {Stiavelli}, {Morelli},
  {Corsini}, {Dalla Bont{\`a}}, {Bradley}  \& {Koekemoer}}{{Calvi}
  et~al.}{2013}]{2013MNRAS.432.3474C}
{Calvi} V.,  {Pizzella} A.,  {Stiavelli} M.,  {Morelli} L.,  {Corsini} E.~M.,
  {Dalla Bont{\`a}} E.,  {Bradley} L.,   {Koekemoer} A.~M.,  2013, \mn@doi
  [\mnras] {10.1093/mnras/stt698}, \href
  {https://ui.adsabs.harvard.edu/abs/2013MNRAS.432.3474C} {432, 3474}

\bibitem[\protect\citeauthoryear{{Choi}, {Hartley}  \& {Dark Energy Survey
  Collaboration}}{{Choi} et~al.}{2020}]{deepfields}
{Choi} A.,  {Hartley} W.,   {Dark Energy Survey Collaboration} 2020, in
  preparation

\bibitem[\protect\citeauthoryear{{Drlica-Wagner} et~al.,}{{Drlica-Wagner}
  et~al.}{2018}]{2018ApJS..235...33D}
{Drlica-Wagner} A.,  et~al., 2018, \mn@doi [\apjs] {10.3847/1538-4365/aab4f5},
  \href {https://ui.adsabs.harvard.edu/abs/2018ApJS..235...33D} {235, 33}

\bibitem[\protect\citeauthoryear{{Euclid Collaboration} et~al.,}{{Euclid
  Collaboration} et~al.}{2019}]{2019A&A...627A..59E}
{Euclid Collaboration} et~al., 2019, \mn@doi [\aap]
  {10.1051/0004-6361/201935187}, \href
  {https://ui.adsabs.harvard.edu/abs/2019A&A...627A..59E} {627, A59}

\bibitem[\protect\citeauthoryear{{Everett}, {Yanny}, {Kuropatkin}  \& {Dark
  Energy Survey Collaboration}}{{Everett} et~al.}{2020}]{balrog2}
{Everett} S.,  {Yanny} B.,  {Kuropatkin} N.,   {Dark Energy Survey
  Collaboration} 2020, in preparation

\bibitem[\protect\citeauthoryear{{Hoekstra}, {Viola}  \&
  {Herbonnet}}{{Hoekstra} et~al.}{2017}]{2017MNRAS.468.3295H}
{Hoekstra} H.,  {Viola} M.,   {Herbonnet} R.,  2017, \mn@doi [\mnras]
  {10.1093/mnras/stx724}, \href
  {https://ui.adsabs.harvard.edu/abs/2017MNRAS.468.3295H} {468, 3295}

\bibitem[\protect\citeauthoryear{{Holwerda}, {Gonzalez}, {Allen}  \& {van der
  Kruit}}{{Holwerda} et~al.}{2005}]{2005AJ....129.1396H}
{Holwerda} B.~W.,  {Gonzalez} R.~A.,  {Allen} R.~J.,   {van der Kruit} P.~C.,
  2005, \mn@doi [\aj] {10.1086/427716}, \href
  {https://ui.adsabs.harvard.edu/abs/2005AJ....129.1396H} {129, 1396}

\bibitem[\protect\citeauthoryear{{Huff} \& {Mandelbaum}}{{Huff} \&
  {Mandelbaum}}{2017}]{metacal1}
{Huff} E.,  {Mandelbaum} R.,  2017, arXiv e-prints, \href
  {https://ui.adsabs.harvard.edu/abs/2017arXiv170202600H} {p. arXiv:1702.02600}

\bibitem[\protect\citeauthoryear{{Jarvis} et~al.,}{{Jarvis}
  et~al.}{2016}]{2016MNRAS.460.2245J}
{Jarvis} M.,  et~al., 2016, \mn@doi [\mnras] {10.1093/mnras/stw990}, \href
  {https://ui.adsabs.harvard.edu/abs/2016MNRAS.460.2245J} {460, 2245}

\bibitem[\protect\citeauthoryear{{Kaiser}, {Squires}  \& {Broadhurst}}{{Kaiser}
  et~al.}{1995}]{1995ApJ...449..460K}
{Kaiser} N.,  {Squires} G.,   {Broadhurst} T.,  1995, \mn@doi [\apj]
  {10.1086/176071}, \href {http://adsabs.harvard.edu/abs/1995ApJ...449..460K}
  {449, 460}

\bibitem[\protect\citeauthoryear{{Kashlinsky}, {Arendt}, {Mather}  \&
  {Moseley}}{{Kashlinsky} et~al.}{2005}]{2005Natur.438...45K}
{Kashlinsky} A.,  {Arendt} R.~G.,  {Mather} J.,   {Moseley} S.~H.,  2005,
  \mn@doi [\nat] {10.1038/nature04143}, \href
  {https://ui.adsabs.harvard.edu/abs/2005Natur.438...45K} {438, 45}

\bibitem[\protect\citeauthoryear{{Kashlinsky}, {Arendt}, {Ashby}, {Fazio},
  {Mather}  \& {Moseley}}{{Kashlinsky} et~al.}{2012}]{2012ApJ...753...63K}
{Kashlinsky} A.,  {Arendt} R.~G.,  {Ashby} M.~L.~N.,  {Fazio} G.~G.,  {Mather}
  J.,   {Moseley} S.~H.,  2012, \mn@doi [\apj] {10.1088/0004-637X/753/1/63},
  \href {https://ui.adsabs.harvard.edu/abs/2012ApJ...753...63K} {753, 63}

\bibitem[\protect\citeauthoryear{{Morganson} et~al.,}{{Morganson}
  et~al.}{2018}]{2018PASP..130g4501M}
{Morganson} E.,  et~al., 2018, \mn@doi [\pasp] {10.1088/1538-3873/aab4ef},
  \href {https://ui.adsabs.harvard.edu/abs/2018PASP..130g4501M} {130, 074501}

\bibitem[\protect\citeauthoryear{{Rowe} et~al.,}{{Rowe}
  et~al.}{2015}]{2015A&C....10..121R}
{Rowe} B.~T.~P.,  et~al., 2015, \mn@doi [Astronomy and Computing]
  {10.1016/j.ascom.2015.02.002}, \href
  {http://adsabs.harvard.edu/abs/2015A%26C....10..121R} {10, 121}

\bibitem[\protect\citeauthoryear{{Samuroff} et~al.,}{{Samuroff}
  et~al.}{2018}]{2018MNRAS.475.4524S}
{Samuroff} S.,  et~al., 2018, \mn@doi [\mnras] {10.1093/mnras/stx3282}, \href
  {https://ui.adsabs.harvard.edu/abs/2018MNRAS.475.4524S} {475, 4524}

\bibitem[\protect\citeauthoryear{{Sheldon} \& {Huff}}{{Sheldon} \&
  {Huff}}{2017}]{metacal2}
{Sheldon} E.~S.,  {Huff} E.~M.,  2017, \mn@doi [\apj]
  {10.3847/1538-4357/aa704b}, \href
  {https://ui.adsabs.harvard.edu/abs/2017ApJ...841...24S} {841, 24}

\bibitem[\protect\citeauthoryear{{Sheldon}, {Becker}, {MacCrann}  \&
  {Jarvis}}{{Sheldon} et~al.}{2019a}]{2019arXiv191102505S}
{Sheldon} E.~S.,  {Becker} M.~R.,  {MacCrann} N.,   {Jarvis} M.,  2019a, arXiv
  e-prints, \href {https://ui.adsabs.harvard.edu/abs/2019arXiv191102505S} {p.
  arXiv:1911.02505}

\bibitem[\protect\citeauthoryear{{Sheldon}, {Becker}, {MacCrann}  \&
  {Jarvis}}{{Sheldon} et~al.}{2019b}]{metadetect}
{Sheldon} E.~S.,  {Becker} M.~R.,  {MacCrann} N.,   {Jarvis} M.,  2019b, arXiv
  e-prints, \href {https://ui.adsabs.harvard.edu/abs/2019arXiv191102505S} {p.
  arXiv:1911.02505}

\bibitem[\protect\citeauthoryear{{Suchyta} et~al.,}{{Suchyta}
  et~al.}{2016}]{2016MNRAS.457..786S}
{Suchyta} E.,  et~al., 2016, \mn@doi [\mnras] {10.1093/mnras/stv2953}, \href
  {https://ui.adsabs.harvard.edu/abs/2016MNRAS.457..786S} {457, 786}

\bibitem[\protect\citeauthoryear{{Tewes}, {Kuntzer}, {Nakajima}, {Courbin},
  {Hildebrandt}  \& {Schrabback}}{{Tewes} et~al.}{2019}]{2019A&A...621A..36T}
{Tewes} M.,  {Kuntzer} T.,  {Nakajima} R.,  {Courbin} F.,  {Hildebrandt} H.,
  {Schrabback} T.,  2019, \mn@doi [\aap] {10.1051/0004-6361/201833775}, \href
  {https://ui.adsabs.harvard.edu/abs/2019A&A...621A..36T} {621, A36}

\bibitem[\protect\citeauthoryear{{Tonry} \& {Schneider}}{{Tonry} \&
  {Schneider}}{1988}]{1988AJ.....96..807T}
{Tonry} J.,  {Schneider} D.~P.,  1988, \mn@doi [\aj] {10.1086/114847}, \href
  {https://ui.adsabs.harvard.edu/abs/1988AJ.....96..807T} {96, 807}

\bibitem[\protect\citeauthoryear{{Verstappen} et~al.,}{{Verstappen}
  et~al.}{2013}]{2013A&A...556A..54V}
{Verstappen} J.,  et~al., 2013, \mn@doi [\aap] {10.1051/0004-6361/201220733},
  \href {https://ui.adsabs.harvard.edu/abs/2013A&A...556A..54V} {556, A54}

\bibitem[\protect\citeauthoryear{{Zuntz}, {Kacprzak}, {Voigt}, {Hirsch}, {Rowe}
   \& {Bridle}}{{Zuntz} et~al.}{2013}]{2013MNRAS.434.1604Z}
{Zuntz} J.,  {Kacprzak} T.,  {Voigt} L.,  {Hirsch} M.,  {Rowe} B.,   {Bridle}
  S.,  2013, \mn@doi [\mnras] {10.1093/mnras/stt1125}, \href
  {https://ui.adsabs.harvard.edu/abs/2013MNRAS.434.1604Z} {434, 1604}

\bibitem[\protect\citeauthoryear{{Zuntz} et~al.,}{{Zuntz} et~al.}{2018}]{y1wl}
{Zuntz} J.,  et~al., 2018, \mn@doi [\mnras] {10.1093/mnras/sty2219}, \href
  {https://ui.adsabs.harvard.edu/abs/2018MNRAS.481.1149Z} {481, 1149}

\makeatother
\end{thebibliography}



\appendix

\section{BFD Validation Simulations}

Shear bias is typically quantified using a linear model $g_{\rm meas} = g_{\rm true}(1+m) + c$. The validation simulations of \citet{2016MNRAS.459.4467B} showed a small but significant $m$ bias ($m = 0.002$) that was unexpected given the underlying design of BFD to be unbiased at small shear values. We have investigated the cause of the bias by re-performing these simulations, varying different properties to identify the cause of the bias.

\begin{table}
	\centering
	\caption{Summary of Simulation Parameters}
	\label{tb:simparams}
\begin{tabular}{ll}
\hline
Characteristic & Parameter \\
 \hline
Galaxy profile & bulge+disk (decentered) \\
Galaxy bulge fraction & $U$(0,1) \\
Galaxy $r_e$ & $U$(1.5-3) pixels \\
Galaxy S/N & $U$(5-25) \\
Galaxy $e$ & $P(e)$ given by equation \ref{eq:pofe} \\
Galaxy shape noise $\sigma_e$ & 0.2 \\
PSF profile & Moffat ($\beta$ = 3.5) \\
PSF size & $r_e$ = 1.5 pixels \\
PSF ellipticity & $e_1$=0.0, $e_2$ = 0.05 \\
Weight function & k$\sigma$ (equation \ref{eq:ksigma}) \\
Weight function $N$ & 4 \\
Weight function $\sigma$ & 3.5 \\
input shear $g_{1,true}$ & [0.02, 0.04, 0.06] \\
input shear $g_{2,true}$ & [0.0, 0.0, 0.0] \\
$N_{batch}$ [target, template] & [5$\times10^5$, 10$^4$] \\
$N_{total}$ & 10$^9$ \\
\hline
\end{tabular}
\end{table}

For the base simulation, we use the \textsc{GalSim} software package \citep{2015A&C....10..121R} to draw galaxies onto postage stamps of size 48x48. Table \ref{tb:simparams} summarizes the basic properties of the simulations, which are similar to those in \citet{2016MNRAS.459.4467B}. The pixel scale is 1\arcsec{} = 1 pixel, and henceforth any angular sizes will be in these units. Each galaxy is composed of a disk and bulge component, where the bulge fraction is drawn from a uniform distribution. The two components are drawn at slightly different offsets from the center of the postage stamp. The half light radius $r_e$ is drawn from a uniform distribution of 1--2 times $r_e$ of the PSF, which is drawn as a Moffat with $r_e$ = 1.5 and $\beta$=3.5. We additionally apply a small ellipticity $e_2$=0.05 to the PSF, to look for any additive bias from imperfect PSF correction. The galaxies are given unlensed ellipticity $e$ = $(a^2 - b^2)/(a^2+b^2)$ components $e_1$ and $e_2$ drawn from the distribution given in equation \ref{eq:pofe} where $\sigma_e$ = 0.2.

\begin{equation}
\label{eq:pofe}
P(e) \propto (1-e^2)^2 \exp{\frac{-e^2}{2\sigma_e^2}}
\end{equation}

The background noise is kept at a constant level for each postage stamp. Galaxies are drawn with flux corresponding to a uniform distribution in S/N over the range of 5-25 for a standard circular galaxy. A constant shear is applied to all galaxies in a given simulation. We use the k$\sigma$ weight function defined in \citet{2016MNRAS.459.4467B} and given in equation \ref{eq:ksigma} with $N$ = 4 and $\sigma$ = 3.5.

\begin{equation}
\label{eq:ksigma}
 W(|k^2|) =  
  \begin{cases}
   (1-\frac{k^2\sigma^2}{2N})^N & k <  \frac{\sqrt{2N}}{\sigma} \\
      0 & k \ge  \frac{\sqrt{2N}}{\sigma}
    \end{cases}
\end{equation}

To speed up tests,  we no longer save images of galaxies to disk, now just saving the BFD moments directly to a table. Each simulation requires 1 billion galaxies to obtain an uncertainty on the multiplicative bias within the $10^{-3}$ multiplicative bias goal. The 1 billion galaxies are divided into batches of 500,000 targets and 10,000 templates. Each batch takes 10 CPU hours, for a total of $\sim20,000$ CPU-hours for an entire simulation. We have used the NERSC computing resources to perform these simulations. 

We have investigated the source of this bias by altering the galaxy population in the simulations (e.g., using a single exponential profile) and by altering the image properties (e.g., the postage stamp size). We determined that the bias arises due to the numerical calculation of the moments and their derivatives. The moments are calculated by summing the Fourier space image of the galaxy's pixel image as given in equation \ref{eq:moments}. The derivatives are also computed as sums over Fourier space, as detailed in Equations C12 and C13 of \citet{2016MNRAS.459.4467B}. We use fast Fourier transforms to convert real space images to Fourier space. The spacing in $k$-space is inversely related to the size of the postage stamp. By using a larger postage stamp size, we create a Fourier space image that is closer to the continuous function to be integrated.

\begin{figure}
    \includegraphics[width=\columnwidth]{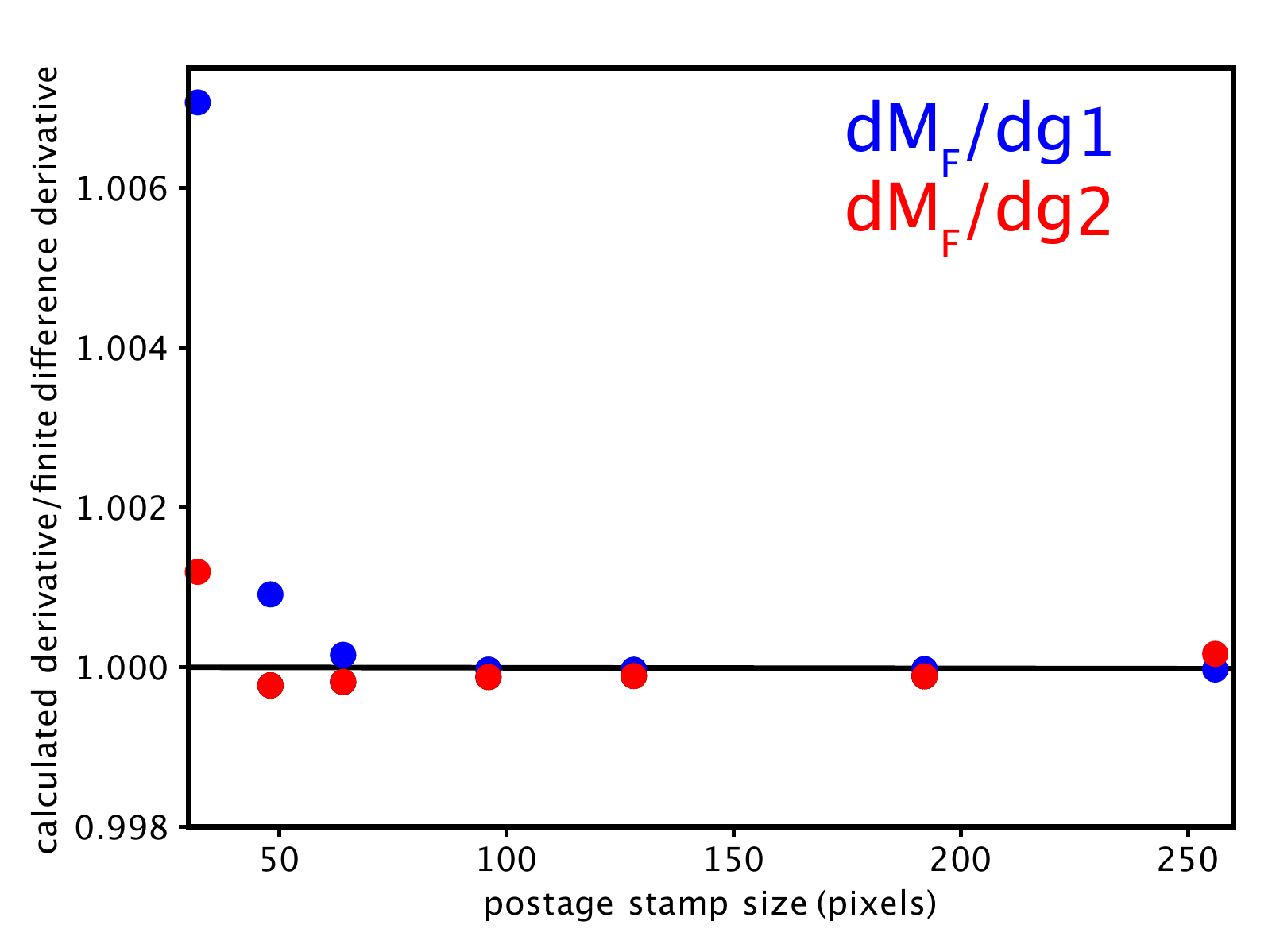}
    \caption{The flux moment derivative with respect to shear computed directly from the unsheared images using the formulas in \citet{2016MNRAS.459.4467B} divided by the same derivative computed using finite differences. The derivatives with respect to $g_1$ and $g_2$ are shown in blue and red respectively. For smaller postage stamp sizes, the difference between the two is large compared to our target of $<10^{-3}$ errors. For our particular setup, a postage stamp size of 96 is sufficient to mitigate the issue. This pattern is seen with other moments and for higher-order derivatives.}
    \label{fig:stampsize}
\end{figure}

We find that the size of the postage stamp is critical for obtaining the correct value of the moments and their derivatives, especially for the $\partial M_F/\partial g_1$.  We compared the derivatives under shear inferred from the unsheared postage stamps to those computed using finite differences between sheared postage stamps.
Figure \ref{fig:stampsize} shows that increasing the postage stamp size results in more precise agreement with the finite-difference estimate.

\begin{figure}
    \includegraphics[width=\columnwidth]{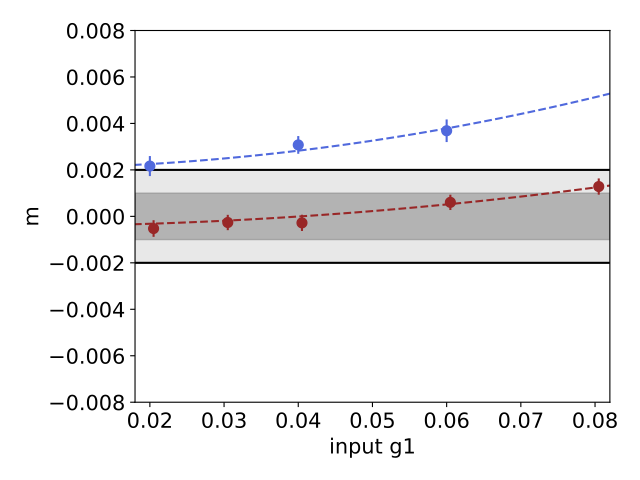}
    \caption{Multiplicative bias $m$ vs. input $g1$ shear. Blue points show data from \citet{2016MNRAS.459.4467B} simulations and red points show data from new simulations using images zero-padded to 96x96. A quadratic fit to the data yields $m = -0.0004$, well within the targets of next generation surveys (grey region).}
    \label{fig:validationsims}
\end{figure}

In real applications, we do not wish to use a large postage stamp, since that increases the interference from neighboring galaxies. Thus, in our simulations, we zero-pad the postage stamps from size $48\times48$ to $96\times96$ before Fourier transforming. In Figure \ref{fig:validationsims}, we show the simulation results from \citet{2016MNRAS.459.4467B}  (blue) and from our new zero-padded validation simulations (red). We find that $m = -0.00052 \pm 0.00035$ for input $g = 0.02$ after zero-padding the images, compared to $m = 0.002 \pm 0.0004$ found in the \citet{2016MNRAS.459.4467B} simulations.  The zero-padding thus yields consistency with $m=0$ and the desired level.

\clearpage
\section*{Affiliations}
\parbox{\textwidth}{
$^{1}$ Department of Physics and Astronomy, University of Pennsylvania, Philadelphia, PA 19104, USA\\
$^{2}$ Department of Physics, ETH Zurich, Wolfgang-Pauli-Strasse 16, CH-8093 Zurich, Switzerland\\
$^{3}$ Kavli Institute for Particle Astrophysics \& Cosmology, P. O. Box 2450, Stanford University, Stanford, CA 94305, USA\\
$^{4}$ Center for Cosmology and Astro-Particle Physics, The Ohio State University, Columbus, OH 43210, USA\\
$^{5}$ Santa Cruz Institute for Particle Physics, Santa Cruz, CA 95064, USA\\
$^{6}$ Department of Physics, Stanford University, 382 Via Pueblo Mall, Stanford, CA 94305, USA\\
$^{7}$ SLAC National Accelerator Laboratory, Menlo Park, CA 94025, USA\\
$^{8}$ Department of Astronomy, University of Illinois at Urbana-Champaign, 1002 W. Green Street, Urbana, IL 61801, USA\\
$^{9}$ National Center for Supercomputing Applications, 1205 West Clark St., Urbana, IL 61801, USA\\
$^{10}$ Jet Propulsion Laboratory, California Institute of Technology, 4800 Oak Grove Dr., Pasadena, CA 91109, USA\\
$^{11}$ Fermi National Accelerator Laboratory, P. O. Box 500, Batavia, IL 60510, USA\\
$^{12}$ Brookhaven National Laboratory, Bldg 510, Upton, NY 11973, USA\\
$^{13}$ Cerro Tololo Inter-American Observatory, National Optical Astronomy Observatory, Casilla 603, La Serena, Chile\\
$^{14}$ Departamento de F\'isica Matem\'atica, Instituto de F\'isica, Universidade de S\~ao Paulo, CP 66318, S\~ao Paulo, SP, 05314-970, Brazil\\
$^{15}$ Laborat\'orio Interinstitucional de e-Astronomia - LIneA, Rua Gal. Jos\'e Cristino 77, Rio de Janeiro, RJ - 20921-400, Brazil\\
$^{16}$ Instituto de Fisica Teorica UAM/CSIC, Universidad Autonoma de Madrid, 28049 Madrid, Spain\\
$^{17}$ LSST, 933 North Cherry Avenue, Tucson, AZ 85721, USA\\
$^{18}$ Physics Department, 2320 Chamberlin Hall, University of Wisconsin-Madison, 1150 University Avenue Madison, WI  53706-1390\\
$^{19}$ Department of Physics \& Astronomy, University College London, Gower Street, London, WC1E 6BT, UK\\
$^{20}$ \\
$^{21}$ Institut de F\'{\i}sica d'Altes Energies (IFAE), The Barcelona Institute of Science and Technology, Campus UAB, 08193 Bellaterra (Barcelona) Spain\\
$^{22}$ INAF-Osservatorio Astronomico di Trieste, via G. B. Tiepolo 11, I-34143 Trieste, Italy\\
$^{23}$ Institute for Fundamental Physics of the Universe, Via Beirut 2, 34014 Trieste, Italy\\
$^{24}$ Observat\'orio Nacional, Rua Gal. Jos\'e Cristino 77, Rio de Janeiro, RJ - 20921-400, Brazil\\
$^{25}$ Centro de Investigaciones Energ\'eticas, Medioambientales y Tecnol\'ogicas (CIEMAT), Madrid, Spain\\
$^{26}$ Department of Physics, IIT Hyderabad, Kandi, Telangana 502285, India\\
$^{27}$ Faculty of Physics, Ludwig-Maximilians-Universit\"at, Scheinerstr. 1, 81679 Munich, Germany\\
$^{28}$ Department of Astronomy/Steward Observatory, University of Arizona, 933 North Cherry Avenue, Tucson, AZ 85721-0065, USA\\
$^{29}$ Department of Astronomy, University of Michigan, Ann Arbor, MI 48109, USA\\
$^{30}$ Department of Physics, University of Michigan, Ann Arbor, MI 48109, USA\\
$^{31}$ Kavli Institute for Cosmological Physics, University of Chicago, Chicago, IL 60637, USA\\
$^{32}$ Institut d'Estudis Espacials de Catalunya (IEEC), 08034 Barcelona, Spain\\
$^{33}$ Institute of Space Sciences (ICE, CSIC),  Campus UAB, Carrer de Can Magrans, s/n,  08193 Barcelona, Spain\\
$^{34}$ Department of Physics, The Ohio State University, Columbus, OH 43210, USA\\
$^{35}$ Center for Astrophysics $\vert$ Harvard \& Smithsonian, 60 Garden Street, Cambridge, MA 02138, USA\\
$^{36}$ Australian Astronomical Optics, Macquarie University, North Ryde, NSW 2113, Australia\\
$^{37}$ Lowell Observatory, 1400 Mars Hill Rd, Flagstaff, AZ 86001, USA\\
$^{38}$ George P. and Cynthia Woods Mitchell Institute for Fundamental Physics and Astronomy, and Department of Physics and Astronomy, Texas A\&M University, College Station, TX 77843,  USA\\
$^{39}$ Department of Astrophysical Sciences, Princeton University, Peyton Hall, Princeton, NJ 08544, USA\\
$^{40}$ Instituci\'o Catalana de Recerca i Estudis Avan\c{c}ats, E-08010 Barcelona, Spain\\
$^{41}$ Department of Physics and Astronomy, Pevensey Building, University of Sussex, Brighton, BN1 9QH, UK\\
$^{42}$ School of Physics and Astronomy, University of Southampton,  Southampton, SO17 1BJ, UK\\
$^{43}$ Computer Science and Mathematics Division, Oak Ridge National Laboratory, Oak Ridge, TN 37831\\
$^{44}$ Institute of Cosmology and Gravitation, University of Portsmouth, Portsmouth, PO1 3FX, UK\\
$^{45}$ Max Planck Institute for Extraterrestrial Physics, Giessenbachstrasse, 85748 Garching, Germany\\
$^{46}$ Universit\"ats-Sternwarte, Fakult\"at f\"ur Physik, Ludwig-Maximilians Universit\"at M\"unchen, Scheinerstr. 1, 81679 M\"unchen, Germany\\
$^{47}$ Institute for Astronomy, University of Edinburgh, Edinburgh EH9 3HJ, UK\\
$^{48}$ D\'{e}partement de Physique Th\'{e}orique and Center for Astroparticle Physics, Universit\'{e} de Gen\`{e}ve, 24 quai Ernest Ansermet, CH-1211 Geneva, Switzerland\\
$^{49}$ Institute of Astronomy, University of Cambridge, Madingley Road, Cambridge CB3 0HA, UK\\
$^{50}$ Brandeis University, Physics Department, 415 South Street, Waltham MA 02453\\
}

\bsp	
\label{lastpage}
\end{document}